\documentclass[final,5p,times]{elsarticle}
\usepackage{amsfonts}
\usepackage{eurosym}
\usepackage{amssymb,bbold}
\usepackage{amsmath}
\usepackage{natbib}
\usepackage{mathrsfs}
\usepackage{hyperref} 
\usepackage{array} 
\usepackage{relsize} 
\usepackage{placeins}
\usepackage{makecell}
\usepackage{ragged2e}
\usepackage{tocloft}
\usepackage{nomencl} 
\usepackage{enumitem}
\usepackage{ulem}
\usepackage{soul}
\usepackage[english]{babel} 

\setlist[itemize,1]{label=$\bullet$} 
\setlist[itemize,2]{label=$\circ$}   
\setlist[itemize,3]{label=$\ast$}    
\setlist[enumerate,1]{label=\arabic*.}  
\setlist[enumerate,2]{label=\alph*),left=1.5em} 
\setlist[enumerate,3]{label=\roman*),left=3.0em} 

\newcolumntype{C}[1]{>{\centering\arraybackslash}m{#1}}
\newcolumntype{L}[1]{>{\centering\arraybackslash}p{#1}}

\makenomenclature

\RequirePackage{color}
\hypersetup{colorlinks=true,linkcolor=blue,citecolor=blue,filecolor=blue,urlcolor=blue,breaklinks=true}
\biboptions{sort&compress}
\begin{document}

\title{Innovative Designs and Insights into Quantum Thermal Machines
}
 
\author[]{Aline D. Lucio}
\ead{aduarte.lucio@ufla.br}
\author[]{Cleverson Filgueiras}
\ead{cleverson.filgueiras@dfi.ufla.br}
\author[]{Moises Rojas}
\ead{moises.leyva@ufla.br}
 
\address{Departamento de F\'{i}sica, Universidade Federal de Lavras, Caixa Postal 3037, 37200-000, Lavras, Minas Gerais, Brazil}

\begin{frontmatter}
\begin{abstract}

We present a comprehensive theoretical investigation about the operational regions of quantum systems, specifically examining their roles as working media functioning between two thermal reservoirs in Quantum Thermal Machines (\nomenclature{QTM}{\hspace{5pt}Quantum Thermal Machine}QTMs). This study provides relevant and novel insights, including a complete spectrum of QTMs within the operational region of these quantum systems, and introduces new QTM designs never before described in the literature. Additionally, this work introduces a standardized and cohesive classification scheme for QTMs, ensuring robustness in nomenclature and operational distinctions, which enhances both theoretical understanding and practical application. Notably, one of these designs directly addresses the need for a more appropriate explanation of the operation of a laser (or maser) as a QTM. Initial calculations were performed to achieve results applicable to any quantum system subjected to rules analogous to those used in classical thermal machine studies. These results were then used to analyze two-level quantum systems as the working medium of QTMs in the Otton cycle. In particular, we analyzed two specific quantum systems: the laser and a spinless electron in a one-dimensional quantum ring, yielding consistent and innovative results. Overall, this study offers valuable insights into the operation and classification of QTMs, establishing a clear and unified framework for their nomenclature while opening new avenues for the design and enhancement of these devices.

\end{abstract}
 
 
\end{frontmatter}

\printnomenclature

\section{Introduction}

The study of quantum thermodynamics has emerged as a vibrant field, driven by both theoretical advancements and experimental progress in manipulating various quantum systems and developing new tools, techniques, and platforms \cite{Springer2018,Dann2023,Adlam2022,Aw2021,Goold2021}. \sloppy{While quantum thermodynamics primarily focuses on non-equilibrium quantum systems \cite{Trushechkin2022},} its significance in studying equilibrium conditions remains crucial for several reasons: establishing foundations and reference points, setting limits, connecting with statistical mechanics, and accessing experimental data \cite{CleversonRef}.

In this context, QTMs have recently attracted significant attention due to their potential to surpass classical limits and enable novel applications in energy conversion and information processing \cite{Myers2022}. Most theoretical studies on QTMs have concentrated on the thermodynamic properties of the Quantum Thermal Engine (\nomenclature{QEN}{\hspace{5pt}Quantum Thermal Engine}QEN) and the Quantum Refrigerator (\nomenclature{QRE}{\hspace{5pt}Quantum Refrigerator}QRE), often comparing them to their classical counterparts \cite{Scovil1959,Kieu2004,Quan2007,Gelbwaser2018,Zheng2014,Torrontegui2017,Uzdin2014,Huang2013,Wang2007,Lin2003,Filgueiras2019,Khlifi2020,ElHawary2023,Kosloff2017}.

This paper addresses these issues by conducting a comprehensive theoretical analysis of the QEN, QRE, and all other possible QTM configurations for quantum systems employed as working media, operating between two thermal reservoirs (\nomenclature{TR}{\hspace{-5pt}Thermal Reservoir}TRs) at distinct temperatures.

In Sec. \ref{sec:OperatingConfigurations}, we systematically explore the operational regions of these systems and the corresponding QTMs that emerge from their specific configurations. New QTM designs naturally arise from considerations of the energy conservation law and the second law of thermodynamics, exhibiting distinct behaviors and efficiencies compared to other configurations. This framework enables a wide exploration of all potential operational configurations of working media. By exploring these operational configurations in Sec. \ref{sec:Efficiencies}, we provide a detailed understanding of QTM efficiencies and their respective Carnot efficiencies, along with relationships between different operational regions and definitions of intersection points between them.

In Sec. \ref{sec:2LET}, we apply these concepts to a quantum system with one quantum particle transitioning between two energy levels, named as \textit{two-level quantum system}, as our working media. Among these systems, we focus specifically on the laser, presenting it as a newly categorized QTM, and investigate all possible operational regions and QTM designs for a spinless electron in a one-dimensional quantum ring \cite{Pereira2022,Viefers2004}, that we named as $e^-$ \textit{in a quantum ring}. Our findings suggest that the laser operates as a QTM within the same operational region as the QEN, but with distinct characteristics and efficiency. For the $e^-$ in a quantum ring system, we identify a broad range of potential QTMs, along with their respective characteristics and efficiencies, particularly when operating in the Otto cycle.

We believe this work makes a significant contribution to the field of quantum thermodynamics by offering new perspectives into QTM design and performance, and we expect that it will inspire further research and development in this domain.

\section{Operating Configurations for QTMs}
  \label{sec:OperatingConfigurations}

\subsection{General Definition of QTMs: Functions and Energy Exchanges}

A QTM is fundamentally a device that operates in cycles, with a quatum system as its working medium connected to two TRs — one at a higher temperature \( T^{h} \) (denoted as \( TR^{h} \)) and the other at a lower temperature \( T^{l} \) (\( TR^{l} \)), that is \( T^{h} > T^{l} \). Energy is exchanged between \( TR^{h} \), \( TR^{l} \), and the QTM's outside environment, through the working medium. The purpose of QTMs is to promote one specific energy exchange while acquiring energy from at least one source in the process.

\begin{figure*}[ht]
    \centering
    \includegraphics[scale=0.38]{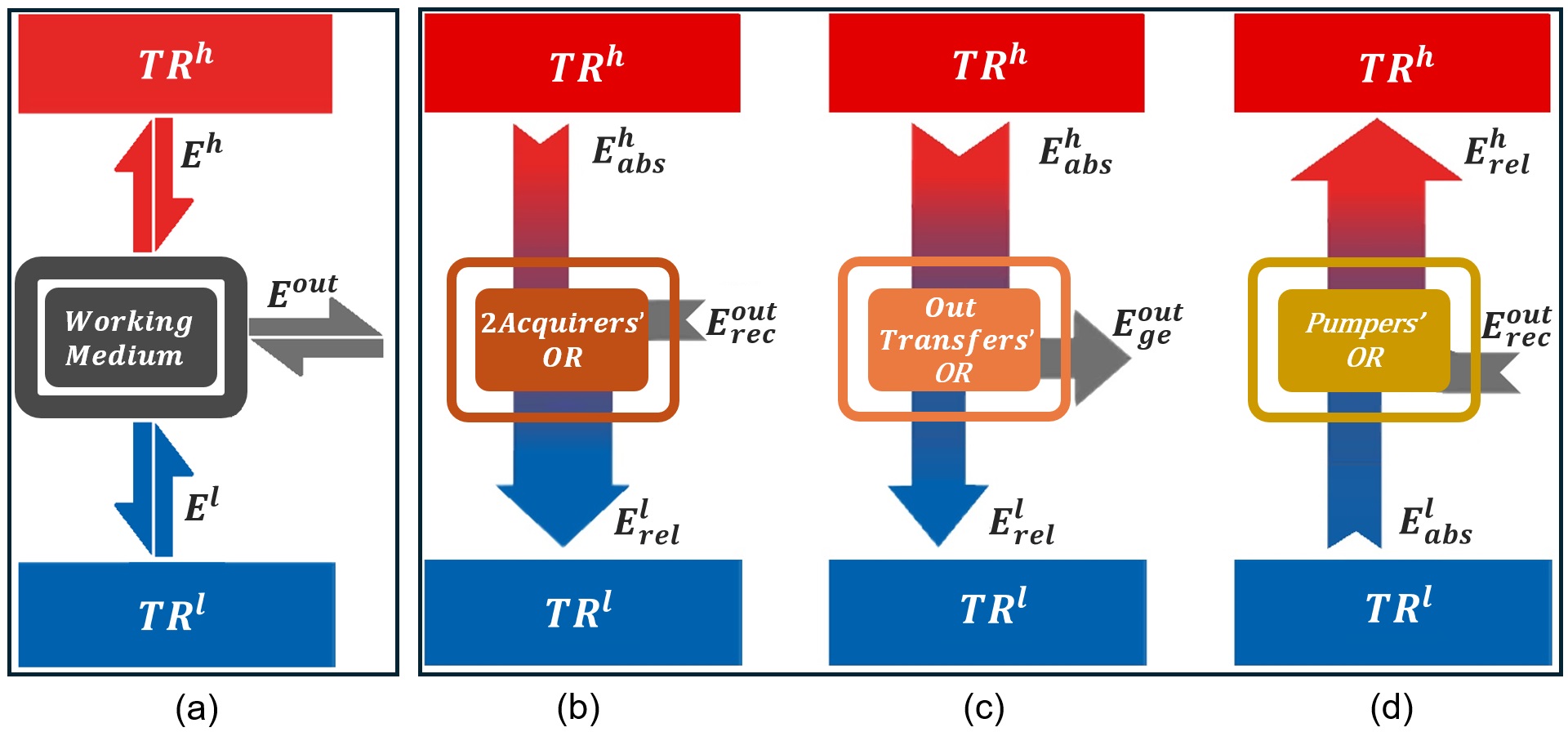}
    \caption{
(a) General schematic of a QTM that consists in a quantum system as the working medium and two reservoirs \( TR^{h} \) and \( TR^{l} \). The function of the working medium is to promote the exchange of energy between the TRs themselves, as well as between the TRs and the QTM's outside environment.  
(b) QTMs operating within the 2Acquirers'OR receive \( E^{out}_{rec} \) from outside, absorb  \( E^{h}_{abs} \) from \( TR^{h} \), and release  \( E^{l}_{rel} \) to \( TR^{l} \). The 2Acquirers'OR includes two subregions: 2Acq$^{out}$, where QTMs utilize \( E^{out}_{rec} \) as their primary energy source, such as QCO and QHT; and 2Acq$^{h}$, where QTMs utilize \( E^{h}_{abs} \) as their primary energy source, such as QDP and QHO. 
(c) QTMs operating within the OutTransfers'OR absorb \( E^{h}_{abs} \) from \( TR^{h} \), generate  \( E^{out}_{ge} \) to outside, and release \( E^{l}_{rel} \) to \( TR^{l} \). This operational region includes QEN and QLL.  
(d) QTMs operating within the Pumpers'OR absorb \( E^{l}_{abs} \) from \( TR^{l} \), receive \( E^{out}_{rec} \) from outside, and release \( E^{h}_{rel} \) to \( TR^{h} \). This operational region includes QRE and QHP.
}
    \label{fig:esquemaQTM}
\end{figure*}

The schematic of a generic QTM is shown in Fig. \ref{fig:esquemaQTM}a. \( E^{h} \) and \( E^{l} \) represent the total energy exchanged with \( TR^{h} \) and \( TR^{l} \), respectively. We introduce the concept of QTM's outside Exchange Energy (denoted as \( E^{out} \)), which represents all energy exchanges by the QTM that are not directly absorbed from or released to \( TR^{h} \) or \( TR^{l} \), i.e., energy exchanged with the QTM's outside beyond the QTM. Specifically, from energy conservation law, we have
\begin{equation}
E^{out}= E^{h} + E^{l}. \label{eq:eq.18}
\end{equation}

Since our calculations are based on the most widely accepted standard in the literature for studying QEN, we define 

\begin{equation}
E^{TR}_{abs} > 0 \ \ \text{and} \ \ E^{TR}_{rel} < 0,
\label{eq:eq.66}
\end{equation}
where \( E^{TR}_{abs} \) refers to the energy absorbed from \( TR^{h} \) or \( TR^{l} \), and \( E^{TR}_{rel} \) refers to the energy released to these TRs. 

In the same way, we define, 
\begin{equation}
 E^{out}_{ge} > 0 \ \ \text{and} \ \ E^{out}_{rec} < 0,
\label{eq:eq.98}
\end{equation}
where \( E^{out}_{ge} \) corresponds to the net generation of energy transferred to the QTM's outside, while the \( E^{out}_{rec} \) signifies the net reception of energy from the QTM's outside. In other words, \( E^{out} \) may originate from or be sent to outside the \( TR^{h} \) - working medium - \( TR^{l} \) axis which is essentially the QTM.

To distinguish the different types of energy exchanges occurring in a QTM cycle, we have standardized the use of specific verbs for each exchange process. Energy acquired from TRs is described as \textit{``absorbed''}, while energy drawn from the QTM's outside is referred to as \textit{``received''}. Conversely, energy directed toward the TRs is termed \textit{``released''}, and energy sent to the QTM's outside is called \textit{``generated''}. This standardization not only facilitates the recognition of each energy exchange event but also helps prevent confusion throughout the analysis. As defined above, energy \textit{``absorbed''} from the TRs or \textit{``generated''} to QTM's outside is treated as positive, whereas energy \textit{``received''} from the QTM's outside or \textit{``released''} to TRs is considered negative.

\subsection{Defining the Operational Regions of a Working Medium and Their Corresponding QTMs}

We define the operational region of a working medium as the set of configurations that allow it to exchange energy with
\( TR^{h} \), \( TR^{l} \), and QTM's outside in a manner consistent with the fundamental principles of thermodynamics. A QTM operates within a specific operational region when its behavior aligns with the defining characteristics of that operational region. The distinction between QTMs in the same operational region is determined by their prioritization of energy exchanges and their primary energy sources.

The general schematic in Fig. \ref{fig:esquemaQTM}a illustrates all possible energy exchanges between the working medium, \( TR^{h} \), \( TR^{l} \), and the QTM's outside.

From this schematic, we identify that only three distinct operational regions emerge, each satisfying the laws of energy conservation and the second law of thermodynamics, while allowing cyclic processes to occur.
 
We refer to the first operational region as the Bi-Acquirers' Operational Region (\nomenclature{2Acquirers'OR}{\hspace{0.16in}Bi-Acquirers' Operational Region}2Acquirers'ORs), where QTMs operate as shown in Fig. \ref{fig:esquemaQTM}b. In this region, \( E^{out}_{rec} \) is received from the QTM's outside, \( E^{h}_{abs} \) is absorbed from \( TR^{h} \), and \( E^{l}_{rel} \) is released to \( TR^{l} \), with \( \left| E^{h}_{abs} \right| < \left| E^{l}_{rel} \right| \). This operational region is unique in having two possible energy sources, resulting in two subregions: 2Acq$^{out}$ and 2Acq$^{h}$. QTMs in 2Acq$^{out}$ subregion use \( E^{out}_{rec} \) as their primary energy source. In this subregion, two types of QTMs are defined: the Quantum Cooler (\nomenclature{QCO}{\hspace{5pt}Quantum Cooler}QCO), which prioritizes the absorption of \( E^{h}_{abs} \) from \( TR^{h} \), and the Quantum Heater (\nomenclature{QHT}{\hspace{5pt}Quantum Heater}QHT), which prioritizes the release of \( E^{l}_{rel} \) to \( TR^{l} \). QTMs in 2Acq$^{h}$ subregion use \( E^{h}_{abs} \) as their primary energy source. In this subregion, two types of QTMs are also identified: the Quantum Thermal Damper (\nomenclature{QDP}{\hspace{5pt}Quantum Thermal Damper}QDP), which prioritizes the reception of \( E^{out}_{rec} \), and the Quantum Heating Optimizer (\nomenclature{QHO}{\hspace{5pt}Quantum Heating Optimizer}QHO), which prioritizes the release of \( E^{l}_{rel} \) to \( TR^{l} \).

Fig. \ref{fig:esquemaQTM}c illustrates the Outside Transfers' Operational Region (\nomenclature{OutTransfers'OR}{\hspace{0.05in}Outside Transfers' Operational Region}OutTransfers'OR). QTMs operating within the OutTransfers'OR absorb \( E^{h}_{abs} \) from \( TR^{h} \) and release \( E^{l}_{rel} \) to \( TR^{l} \), where \( \left| E^{h}_{abs} \right| > \left| E^{l}_{rel} \right| \). In this process, \( E^{out}_{ge} \) is generated. The well-studied QEN, which prioritizes the generation of \( E^{out}_{ge} \), and the newly-designed Quantum Thermal Laser-
Like (\nomenclature{QLL}{\hspace{5pt}Quantum Thermal Laser-Like}QLL), which prioritizes the release of \( E^{l}_{rel} \), are QTMs operating within this region. As its name suggests, we believe that the QLL provides a more accurate representation of the laser functionality as a QTM. (see Subsec. \ref{sec:2LET}-\ref{subsec:Laser}).

Finally, Fig. \ref{fig:esquemaQTM}d illustrates the Thermal Pumpers' Operational Region (\nomenclature{Pumpers'OR}{\hspace{0.29in}Thermal Pumpers' Operational Region}Pumpers'OR). QTMs in this operational region operate in a cycle opposite to that of the QTMs in the OutTransfers'OR. Here, \( E^{h}_{rel} \) is released from \( TR^{h} \), and \(E^{l}_{abs} \) is absorbed from \( TR^{l} \), where \( \left| E^{h}_{rel} \right| > \left| E^{l}_{abs} \right| \). This process does not occur naturally between the TRs and requires the input of \( E^{out}_{rec} \) from the QTM's outside to enable the exchange. It is important to emphasize that \( E^{l}_{abs} \) cannot be considered an energy source, as doing so would contradict the second law of thermodynamics. Two QTMs operate within the Pumpers'OR: the well known QRE, which is essentially a Cold Pumper and prioritizes the absorption of \( E^{l}_{abs} \) from \( TR^{l} \), and 
the Quantum Heat Pumper (\nomenclature{QHP}{\hspace{5pt}Quantum Heat Pumper}QHP), which prioritizes the release of \( E^{h}_{rel} \) to \( TR^{h} \).

It would be preferable to use nomenclatures without terms like \textit{``cold''}, \textit{``heat''} or \textit{``refrigerator''} as these may seem inconsistent with quantum phenomena. However, we have chosen to retain these classical nomenclatures because these terms are widely used in the study of QTMs.

We are dealing with a working medium coupled to two TRs. Consequently, if the working medium receives energy from one TR, it must release energy to the other for the cycle to occur, that is,

 \begin{equation}
    \begin{split}
            \mathlarger{ E^{l}_{abs} \Leftrightarrow E^{h}_{rel}}, \\
             \mathlarger{E^{h}_{abs} \Leftrightarrow E^{l}_{rel}}.
            \label{eq:eq.99}
    \end{split}
\end{equation}

Moreover, since the aim is to present the results for all operational regions of a working medium in a single graph, and given the definition in Eq. \ref{eq:eq.66} and the implications summarized in Eqs. \ref{eq:eq.99}, we can express a common equation between $E^h$ and $E^l$ for all operational regions as follows

\begin{equation}
\mathlarger{\frac{E^{h}}{E^{l}} = -\alpha^{2}.} \label{eq:eq.47}
\end{equation}
where we named $\alpha^{2}$ as the \textit{thermal high-low energy ratio}.

Thus, considering the energy values as defined earlier for any exchange within any operational region, we conclude that \( 0 < \alpha^{2} < 1 \) corresponds to the relationship between \(E^h\) and \(E^l\) in the 2Acquirers'OR, while \( \alpha^{2} > 1 \) represents this relationship in the OutTransfers'OR and Pumpers'OR.

\subsection{Classical and Quantum Energy Relationships}

The first law of thermodynamics for a classical working medium imposes the following: 

\begin{equation}
dU = \delta{Q} - \delta{W}, \label{eq:eq.40}
\end{equation}
which, in quasi-static processes, can be written as

\begin{equation}
\Delta U = Q - W, \label{eq:eq.42}
\end{equation}
where \(\Delta U\) represents the change in the internal energy of the classical working medium, \( Q = Q_{abs} + Q_{rel} \) denotes the total heat exchanged with the TRs, and \( W = W_{ge} + W_{rec} \) refers to the net work resulting from the interaction with the machine's outside. In this case, the classical working medium forms a classical thermal machine in combination with the TRs. According to Eq. \ref{eq:eq.42}, the heat absorbed from the TRs, \( Q_{abs} \), and the work done by the classical working medium on the machine's outside, represented by \( W_{ge} \), are both positive values. Conversely, the heat released to the TRs, \( Q_{rel} \), and the work done on the classical working medium by the machine's outside, represented by \( W_{rec} \), are negative values.

Since the operation of a classical thermal machine implies that the thermodynamic processes are cyclic, that is, \(\Delta U = 0\), Eq. \ref{eq:eq.42} for this machine will be given by \(W = Q\) or, by making \(Q = Q_{abs} + Q_{rel}\),

\begin{equation}
W = Q_{abs} + Q_{rel}. \label{eq:eq.50}
\end{equation}

The main reason for not substituting \( W = W_{ge} + W_{rec} \) into Eq. \ref{eq:eq.50} is that, for a specific thermal machine, distinguishing between \( W_{ge} \) and \( W_{rec} \) is often unnecessary. Considering the characteristics of classical thermal machines, it is more typical to consider \( W = W_{ge} \)  when the net work is positive and \( W = W_{rec} \) when the net work is negative."

For instance, when evaluating the efficiency of a thermal engine, the critical question is: \textit{``How much useful work can it perform?''} In such cases, in Eq. \ref{eq:eq.50}, \( W \) represents the net or useful work produced by the classical working medium in each engine cycle, specifically \( W = W_{ge} \).

By comparing Eq. \ref{eq:eq.18} with Eq. \ref{eq:eq.50}, and considering Eq.\ref{eq:eq.99}, we can conclude that 
 \begin{equation}
    \begin{split}
            \mathlarger{\left[E^{out}\right]_{\textstyle \text{\tiny QTM}}} &\mathlarger{\equiv  \left[W\right]_{\textstyle \text{\tiny CTM}}},\\
            \mathlarger{\left[ E^{h} + E^{l}\right]_{\textstyle \text{\tiny QTM}}} & \mathlarger{\equiv  \left[Q\right]_{\textstyle \text{\tiny CTM}}}, \text{and}\\
            \mathlarger{\left[ E^{TR}_{abs} + E^{TR}_{rel}\right]_{\textstyle \text{\tiny QTM}}} & \mathlarger{\equiv  \left[Q\right]_{\textstyle \text{\tiny CTM}}},  
            \label{eq:eq.100}
    \end{split}
\end{equation}
where $\text{\small CTM}$ is an acronym for \textit{classical thermal machine}.

The last equivalence above can also be expressed as
 \begin{equation}
    \begin{split}
            \mathlarger{\left[ E^{TR}_{abs} + E^{TR}_{rel}\right]_{\textstyle \text{\tiny QTM}}} & \mathlarger{\equiv  \left[Q_{abs} + Q_{rel}\right]_{\textstyle \text{\tiny CTM}}}, \text{or even}\\
            \mathlarger{\left[ E^{TR}_{abs}\right]_{\textstyle \text{\tiny QTM}}} & \mathlarger{\equiv  \left[Q_{abs}\right]_{\textstyle \text{\tiny CTM}}}, \text{and}\\
             \mathlarger{\left[ E^{TR}_{rel}\right]_{\textstyle \text{\tiny QTM}}} & \mathlarger{\equiv  \left[Q_{rel}\right]_{\textstyle \text{\tiny CTM}}}.
            \label{eq:eq.101}
    \end{split}
\end{equation}

Despite the formal equivalence between work and heat in terms of energy exchanges described in this paper, maintaining our generalized definitions and nomenclature for these quantities is essential to ensure a comprehensive analysis of QTMs.

\section{Efficiency Calculation and Carnot Efficiency}
\label{sec:Efficiencies}

Typically, \textit{``efficiency''} is denoted by the symbol $\eta$ or $\varepsilon$, and \textit{``coefficient of performance''} by the symbol COP, depending on how the QTM works. For convenience, we use ``efficiency'' and the symbol $\varepsilon$ to represent both efficiencies and coefficients of performance, as their distinction will be clear throughout the work, without the need to carry two distinct symbols. Generally, we can write the QTM efficiency, $\varepsilon_{\textstyle \text{\tiny QTM}}$, as
\begin{equation}
\mathlarger{\varepsilon_{\textstyle \text{\tiny QTM}} = \frac{\left| E_{target} \right|}{\left| E_{available} \right|}}, \label{eq:eq.17}
\end{equation}
where $E_{target}$ is the energy that the QTM prioritizes and $E_{available}$ is the main energy source available for the QTM to achieve its objective. 

At the same way, the QTM Carnot efficiency, $\varepsilon_{{\textstyle \text{\tiny QTM}}_c}$, should write as

\begin{figure*}[ht]
    \centering
    \includegraphics[scale=0.28]{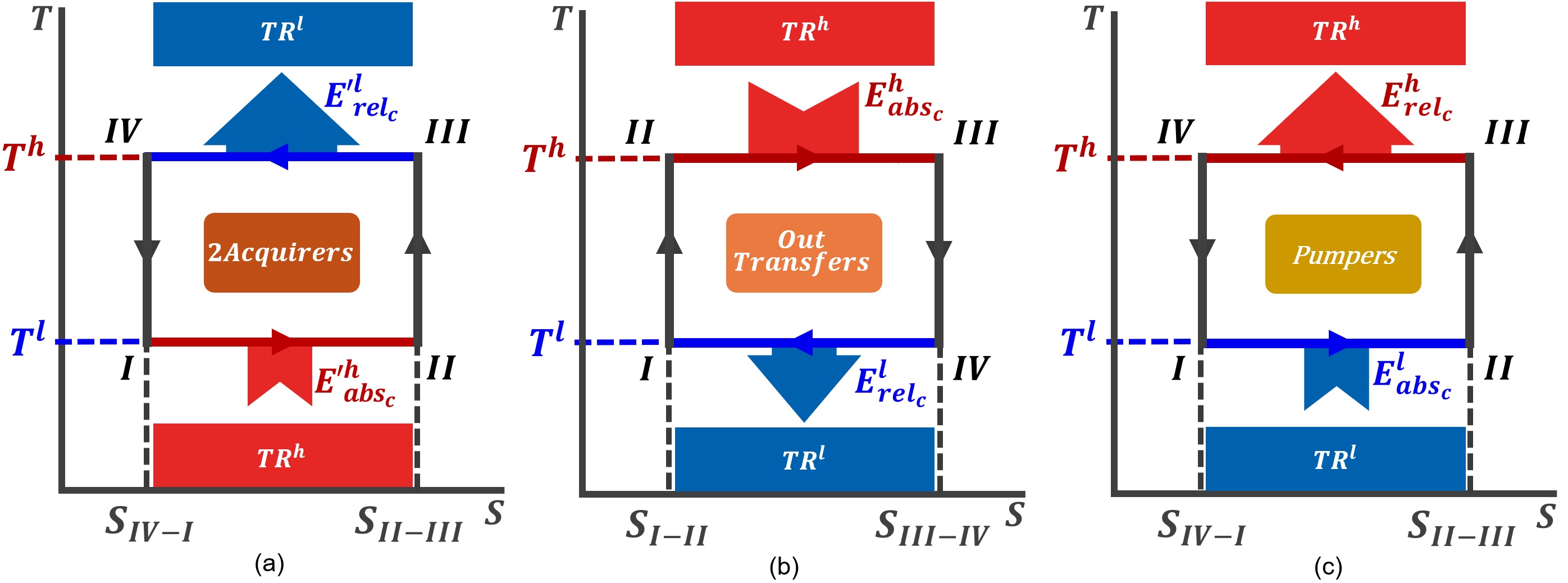}
     \caption{Representation of the Carnot cycle in a T-S diagram for each operational region. The \( I-II-III-IV \) notation 
represents the thermodynamic states of the working medium and the cycle direction. This notation, as shown in the diagrams, serves as the standard for all analyses 
in this study. \( T^{h} \) and \( T^{l} \) represent the maximum and minimum temperature reached by the QTMs during the cycle, corresponding to the 
temperature of \( TR^{h} \) and \( TR^{l} \), respectively. (a) The counterclockwise Carnot cycle diagram for QTMs in the 2Acquirers'OR, where \( S_{IV-I} \) denotes the 
working medium entropy in states IV and I, and \( S_{II-III} \) in states II and III. (b) The clockwise Carnot cycle diagram for QTMs in the OutTransfers'OR, where \( S_{I-II} \) refers to the 
working medium entropy in states I and II, and \( S_{III-IV} \) in states III and IV. (c) The counterclockwise Carnot cycle diagram for QTMs in the Pumpers'OR, where \( S_{IV-I} \) 
indicates the working medium entropy in states IV and I, and \( S_{II-III} \) in states II and III.}
    \label{fig:CCarnotQTM}
\end{figure*}

\begin{equation}
\mathlarger{\varepsilon_{{\textstyle \text{\tiny QTM}}_c}= \frac{\left| E_{{target}_c} \right|}{\left| E_{{available}_c} \right|}}, \label{eq:eq.65}
\end{equation}
where \( E_{{target}_c} \) represents the energy prioritized by the QTM when operating in the Carnot cycle, and \( E_{{available}_c} \) denotes the energy source available for the QTM to achieve its objective within this cycle.

The efficiency definitons \ref{eq:eq.17} and \ref{eq:eq.65} are already well-known and studied. However, with the proposal of division into operational regions, for the working medium and the introduction of new QTMs, a broader understanding of the $\varepsilon_{{\textstyle \text{\tiny QTM}}_c}$ function becomes essential. 

The design of any QTM, where the working medium interacts with both $TR^h$ and $TR^l$, imposes constraints on the values of the exchanged energy. From the diagrams in Fig. \ref{fig:esquemaQTM}, we observe that \(E^l\) and \(E^h\) cannot be made arbitrarily small or large, as the temperatures of the TRs are finite. The upper and lower bounds for these energy values are determined by operating the QTM in the Carnot cycle, which theoretically provides the limits for the $\varepsilon_{\textstyle \text{\tiny QTM}}$, whereas this cycle places the QTM in its most optimal configuration.

Therefore, we define \( \varepsilon_{{\textstyle \text{\tiny QTM}}_c} \) as the limiting efficiency of a QTM operating in any cycle. It is important to note that this definition is comprehensive, as this limit may correspond to either a maximum or minimum value of \( \varepsilon_{\textstyle \text{\tiny QTM}} \), whose the significance become clear in later analyses.

\subsection{Carnot Cycle}
\label{subsec: Carnot Cycle}

Figure \ref{fig:CCarnotQTM} illustrates the three operational regions in Temper\-ature-Entropy (T-S) diagrams, representing all potential configurations for a QTM operating in the Carnot cycle. The notation \( I-II-III-IV \) designates the thermodynamic states and the direction of the cycle for each operational region, establishing a standard framework for analysis across the diagrams in this work. \( T^{h} \) and \( T^{l} \) correspond to the maximum and minimum temperatures reached by the QTMs during the cycle, which align with the temperatures of the thermal reservoirs \( TR^{h} \) and \( TR^{l} \), respectively.

For simplifying several aspects of our analysis, we define the \textit{high-low temperature ratio}, \(\theta^2\), as

\begin{equation}
\mathlarger{\theta^{2} = \frac{T^h}{T^l}},
\label{eq:eq.24}
\end{equation} 
where $\theta^{2} > 1$ for any QTM configuration.

In the previous section, we concluded that comparing Eq. \ref{eq:eq.18} with Eq. \ref{eq:eq.50} reveals that the energy exchanges between the working medium and the thermal reservoirs, \( TR^{h} \) and \( TR^{l} \), are analogous to classical heat exchanges. This analogy is well-supported in the literature (see \cite{Myers2022} and \cite{Quan2007}), forming a basis for applying the following classical relationship between entropy (\( S \)), temperature (\( T \)), and heat (\( Q \)), in the study of quantum systems:

\begin{equation}
     \mathlarger{\delta{Q}=Td{S}}.  \label{eq:eq.32}
\end{equation}

Eq. \ref{eq:eq.32} enables the calculation of entropy variation for the working medium in any operational region, assuming $Q$ given by Eqs. \ref{eq:eq.101}.

The QTMs operating within the 2Acquirers'OR follow a counterclockwise Carnot cycle, as shown in Fig. \ref{fig:CCarnotQTM}a. During this process, \( E'^{l}_{{rel}_c} \) is released to \( TR^{l} \) in an isothermal stroke at \( T^{h} \), which exceeds the \( E'^{h}_{{abs}_c} \) absorbed from \( TR^{h} \) during an isothermal stroke at \( T^{l} \). Essentially, the reception of \( E^{out}_{rec} \) from the machine's outside positions the QTM such that the working medium reaches its maximum temperature (\( T^{h} \)) just before releasing energy to \( TR^{l} \).

In the Carnot cycle, the entropy variation of the working medium for QTMs in the 2Acquirers'OR is expressed as

\begin{equation}
  \mathlarger{\Delta{S_{{2Acq}_{Carnot}}}=\int_{I}^{II} \frac{dE'^{h}_{{abs}_c}}{T^l} + \int_{III}^{IV} \frac{dE'^{l}_{{rel}_c}}{T^h}.} \label{eq:eq.02}
\end{equation}

In the strokes $II$-$III$ and $IV$-$I$, there is no entropy change as these are adiabatic processes, and there is no entropy change in a closed cycle. Thus, considering that the strokes $I-II$ and $III-IV$ are quasi-static isothermal processes, from Eq. \ref{eq:eq.24} and Eq. \ref{eq:eq.02}, we have

\begin{equation}
\mathlarger{\frac{T^h}{T^l} = -\frac{E'^{l}_{{rel}_c}}{E'^{h}_{{abs}_c}}= \theta^{2}.} \label{eq:eq.56}
\end{equation}

The QTMs in the OutTransfers'OR in a clockwise cycle (Fig. \ref{fig:CCarnotQTM}b), absorbing the energy $E^{h}_{{abs}_c}$ during an isothermic stroke at $T^h$ and releasing the energy $E^{l}_{{rel}_c}$ during an isothermic stroke at $T^l$. 

Building on the same analysis that led to Eq. \ref{eq:eq.02}, we obtain the following result for the entropy variation of the QTMs' working medium in the OutTransfers'OR

\begin{equation}
  \mathlarger{\Delta{S_{{OutT}_{Carnot}}}=\int_{II}^{III} \frac{dE^{h}_{{abs}_c}}{T^h} + \int_{IV}^{I} \frac{dE^{l}_{{rel}_c}}{T^l}}, \label{eq:eq.21}
\end{equation}
and, from Eq. \ref{eq:eq.24} and Eq. \ref{eq:eq.21}, we have
\begin{equation}
\mathlarger{\frac{T^h}{T^l} = -\frac{E^{h}_{{abs}_c}}{E^{l}_{{rel}_c}}=\theta^{2}.} \label{eq:eq.01}
\end{equation}

Fig. \ref{fig:CCarnotQTM}c shows the counterclockwise cycle for the QTMs in the Pumpers'OR, where $E^{l}_{{abs}_c}$ is absorved during an isothermic stroke at $T^l$ and $E^{h}_{{rel}_c}$ is released during an isothermic stroke at $T^h$. The Carnot cycle follows by these QTMs is the exact opposite to the Carnot cycle followed by the QTMs in the OutTransfers'OR, and the working medium's entropy variation is given by

\begin{equation}
  \mathlarger{\Delta{S_{{Pump}_{Carnot}}}=\int_{I}^{II} \frac{dE^{l}_{{abs}_c}}{T^l} + \int_{III}^{IV} \frac{dE^{h}_{{rel}_c}}{T^h}}, \label{eq:eq.08}
\end{equation}
and

\begin{equation}
\mathlarger{\frac{T^h}{T^l} = -\frac{E^{h}_{{rel}_c}}{E^{l}_{{abs}_c}}=\theta^{2}.} \label{eq:eq.48}
\end{equation}

Next, we apply these results to calculate the efficiency \( \varepsilon_{\textstyle \text{\tiny QTM}} \) and the Carnot efficiency \( \varepsilon_{{\textstyle \text{\tiny QTM}}_c} \) for each type of QTM operating within its respective operational region. These calculations provide insight into how different configurations and energy exchanges impact the overall performance of QTMs.

\subsection{Bi-Acquirers' Operational Region}
In the 2Acquirers'OR, QTMs can use two distinct acquired energies as their primary source: $E^{out}_{rec}$ or $E^{h}_{abs}$. This results in two subregions: 2Acq$^{out}$ and 2Acq$^{h}$. QTMs operating within 2Acq$^{out}$ subregion, such as the QCO and the QHT, use $E^{out}_{rec}$ as the main energy source. Conversely, QTMs in 2Acq$^{h}$ subregion, like the QDP and QHO, use $E^{h}_{abs}$.

Both QCO and QHT have already been referenced by other authors. However, their characterizations and nomenclatures remain inconsistently defined, often used arbitrarily, which obstructs a broader understanding of their actual potential and limitations.

\subsubsection {Quantum Cooler}

\begin{itemize}
  \item {QCO Operational Mode} 
    \begin{itemize}
     \item {Energy priority: $E_{target}=E^{h}_{abs}$}.
  
     \item {Energy source: $E_{available}=E^{out}_{rec}$}. 
    \end{itemize}
    
  \item {QCO Efficiency}  
  
  From Eq. \ref{eq:eq.17}, we can write
 \begin{align}
    \mathlarger{\varepsilon_{\textstyle \text{\tiny QCO}}} &= \mathlarger{\frac{\left| E^{h}_{abs}\right|}{\left| E^{out}_{rec}\right|} = \frac{E^{h}_{abs}}{-\left( E^{h}_{abs}+E^{l}_{rel}\right)} =}\nonumber \\
    &= \mathlarger{\frac{1}{-\frac{E^{l}_{rel}}{E^{h}_{abs}}-1}},  \label{eq:eq.09}
    \end{align}
whose substitutions were based on the Eqs. \ref{eq:eq.18} and \ref{eq:eq.66}.

Using Eq. \ref{eq:eq.47}, in Eq. \ref{eq:eq.09}, we have

 \begin{align}
    \mathlarger{\varepsilon_{\textstyle \text{\tiny QCO}}=\frac{1}{\frac{1}{\alpha^{2}}-1}.}  \label{eq:eq.67}
    \end{align}
 
  \item {\textit{QCO Carnot Efficiency} } 
    
   From Eqs. \ref{eq:eq.18}, \ref{eq:eq.66} and \ref{eq:eq.65}, we can write
    
    \begin{equation}
    \begin{aligned}
    \mathlarger{\varepsilon_{{\textstyle \text{\tiny QCO}}_c}} &= \mathlarger{\frac{\left|E'^{h}_{{abs}_c}\right|}{\left|E'^{E}_{{rec}_c}\right|}=\frac{1}{-\frac{E'^{l}_{{rel}_c}}{E'^{h}_{{abs}_c}}-1}.}
    \end{aligned}
    \label{eq:eq.43}
    \end{equation}

Using Eq. \ref{eq:eq.56} in Eq. \ref{eq:eq.43} we have

 \begin{equation}
 \mathlarger{\varepsilon_{{\textstyle \text{\tiny QCO}}_c}=\frac{1}{\frac{T^h}{T^l}-1}=\frac{1}{\theta^{2}-1}.} \label{eq:eq.07}
 \end{equation} 	  

It is very important to emphasize that $\theta$ it depends exclusively on the ratio between $T^h$ and $T^l$, as given by Eq. \ref{eq:eq.24}.

\item {\textit{Analysis of Limits for $\varepsilon_{\textstyle \text{\tiny QCO}}$ and $\alpha^{2}_{\textstyle \text{\tiny QCO}}$ Values}}

Eq. \ref{eq:eq.47} is defined such that $0 < \alpha^{2} < 1$ represents the relationship between $E^h$ and $E^l$ for any QTM within the 2Acquirers'OR. The implication of this restriction in Eq. \ref{eq:eq.67} is as follows:
    \begin{itemize}
     \item {When $\alpha^{2}  \rightarrow 0$, $\varepsilon_{\textstyle \text{\tiny QCO}} \rightarrow 0$.}

     \item {When $\alpha^{2}  \rightarrow 1$, $\varepsilon_{\textstyle \text{\tiny QCO}} \rightarrow \infty$.}
    \end{itemize}
 There is no impediment to \( \varepsilon_{\textstyle \text{\tiny QCO}} \rightarrow 0 \). Therefore, there are also no restrictions on the minimum value of \(\alpha\), allowing for 
 
\begin{equation}
\mathlarger{\alpha^{2}_{{\textstyle \text{\tiny QCO}}_{min}}=0}. \label{eq:eq.90}
\end{equation}
 
 However, when any $\varepsilon_{{\textstyle \text{\tiny QTM}}} \rightarrow \infty$, it is necessary to impose a limit, which is associated to the ideal capacity of the TRs to supply and absorb energy from the working medium, i.e. 

\begin{equation}
\mathlarger{\varepsilon_{{\textstyle \text{\tiny QCO}}_{max}} = \varepsilon_{{\textstyle \text{\tiny QCO}}_c} \bigg|_{\alpha^{2}  \rightarrow 1},} \label{eq:eq.68}
\end{equation}
where $\alpha^2$ has a maximum value for the QCO, obtained by substituting Eqs. \ref{eq:eq.67} and \ref{eq:eq.07} into the above equation, such that

 \begin{align}
    \mathlarger{\frac{1}{\frac{1}{\alpha^{2}_{{\textstyle \text{\tiny QCO}}_{max}}}-1}=\frac{1}{\theta^{2}-1},} \nonumber 
    & \\ \mathlarger{ \alpha^{2}_{{\textstyle \text{\tiny QCO}}_{max}} = \frac{1}{\theta^{2}}.}
    \label{eq:eq.76}
    \end{align}

\end{itemize}

For all subsequent calculations of $\varepsilon_{\textstyle \text{\tiny QTM}}$ and $\varepsilon_{{\textstyle\text{\tiny QTM}}_c}$, the substitutions from Eqs. \ref{eq:eq.18}, \ref{eq:eq.66}, and \ref{eq:eq.47} through \ref{eq:eq.24} will be applied implicitly. In the specific case of calculating $\varepsilon_{{\textstyle \text{\tiny QTM}}_c}$ within the 2Acquirers'OR, the substitution based on Eq. \ref{eq:eq.56} will also be 
considered implicit.

\subsubsection {Quantum Heater}

\begin{itemize}
  \item {QHT Operational Mode} 
    \begin{itemize}
     \item {Energy priority: $E_{target}=E^{l}_{rel}$}.
  
     \item {Energy source: $E_{available}=E^{out}_{rec}$}.
    \end{itemize}
    
  \item {QHT Efficiency}  
  \begin{align}
    \mathlarger{\varepsilon_{\textstyle \text{\tiny QHT}} = \frac{\left| E^{l}_{rel}\right|}{\left| E^{out}_{rec}\right|}=\frac{1}{1+\frac{E^{h}_{abs}}{E^{l}_{rel}}} = \frac{1}{1-\alpha^{2}}.}  \label{eq:eq.69}
    \end{align}
    
  \item {\textit{QHT Carnot Efficiency} }  
    \begin{equation}
    \begin{aligned}
    \mathlarger{\varepsilon_{{\textstyle \text{\tiny QHT}}_c} = \frac{1}{1+\frac{E'^{h}_{{abs}_c}}{E'^{l}_{{rel}_c}}}=\frac{1}{1-\frac{T^l}{T^h}}=\frac{1}{1-\frac{1}{\theta^{2}}}.}
    \end{aligned}
    \label{eq:eq.70}
    \end{equation}

\item {\textit{Analysis of Limits for $\varepsilon_{\textstyle \text{\tiny QHT}}$ and $\alpha^{2}_{\textstyle \text{\tiny QHT}}$ Values}}

For QHT, $0 < \alpha^{2} < 1$. In Eq. \ref{eq:eq.69} follows that:
    \begin{itemize}
     \item {When $\alpha^{2}  \rightarrow 0$, $\varepsilon_{\textstyle \text{\tiny QHT}} \rightarrow 1$.}

     \item {When $\alpha^{2}  \rightarrow 1$, $\varepsilon_{\textstyle \text{\tiny QHT}} \rightarrow \infty$.}
    \end{itemize}

Since it is allowed for \( \varepsilon_{\textstyle \text{\tiny QHT}} \rightarrow 1 \), we have

\begin{equation}
\mathlarger{\alpha^{2}_{{\textstyle \text{\tiny QHT}}_{min}}=0}. \label{eq:eq.91}
\end{equation}

On the other hand,

\begin{equation}
\mathlarger{\varepsilon_{{\textstyle \text{\tiny QHT}}_{max}} = \varepsilon_{{\textstyle \text{\tiny QHT}}_c} \bigg|_{\alpha^{2}  \rightarrow 1}.} \label{eq:eq.71}
\end{equation}

As derived for the QCO, the value of \(\alpha^2\) reaches a maximum for the QHT, determined by inserting Eqs. \ref{eq:eq.69} and \ref{eq:eq.70} into the Eq. \ref{eq:eq.71}, resulting in

 \begin{align}
    \mathlarger{ \alpha^{2}_{{\textstyle \text{\tiny QHT}}_{max}} = \frac{1}{\theta^{2}}.}
    \label{eq:eq.77}
    \end{align}

\end{itemize}

\subsubsection { Quantum Thermal Damper}

Currently, no mention in the literature addresses a QTM with the specific configuration proposed for the QDP. The main priority of this QTM is to receive $E^{out}_{rec}$ from outside. As we will demonstrate, $\varepsilon_{\textstyle \text{\tiny QDP}}$ is inversely related to that of the QEN 
($\varepsilon_{\textstyle \text{\tiny QEN}}$). While a QEN generates useful $E^{out}_{ge}$ to outside, the QDP prioritizes extracting $E^{out}_{rec}$ from the outside, functioning as a quantum damper.

\begin{itemize}
  \item {QDP Operational Mode} 
    \begin{itemize}
     \item {Energy priority: $E_{target}=E^{out}_{rec}$}.
  
     \item {Energy source: $E_{available}=E^{h}_{abs}$}. 
    \end{itemize}
    
  \item {QDP Efficiency}  
  \begin{align}
    \mathlarger{ \varepsilon_{\textstyle \text{\tiny QDP}} = \frac{\left| E^{out}_{rec}\right|}{\left| E^{h}_{abs}\right|}=-\frac{E^{l}_{rel}}{E^{h}_{abs}}-1 = \frac{1}{\alpha^{2}}-1.}  \label{eq:eq.72}
    \end{align}
    
  \item {\textit{QDP Carnot Efficiency} }  
    \begin{equation}
    \begin{aligned}
    \mathlarger{\varepsilon_{{\textstyle \text{\tiny QDP}}_c} =-\frac{E'^{l}_{rel}}{E'^{h}_{abs}}-1 =\frac{T^h}{T^l}-1=\theta^{2}-1.}
    \end{aligned}
    \label{eq:eq.73}
    \end{equation}

\item {\textit{Analysis of Limits for $\varepsilon_{\textstyle \text{\tiny QDP}}$ and $\alpha^{2}_{\textstyle \text{\tiny QDP}}$ Values}}

For QDP, $0 < \alpha^{2} < 1$. In Eq. \ref{eq:eq.72} follows that:
    \begin{itemize}
     \item {When $\alpha^{2}  \rightarrow 0$, $\varepsilon_{\textstyle \text{\tiny QDP}} \rightarrow \infty$.}

     \item {When $\alpha^{2}  \rightarrow 1$, $\varepsilon_{\textstyle \text{\tiny QDP}} \rightarrow 0$.}
    \end{itemize}

In this case, we observe that \( \varepsilon_{\textstyle \text{\tiny QDP}} \rightarrow 0 \) is also allowed, as well as it is allowed for \(\alpha^2\) to reach its maximum value,i.e.,

\begin{equation}
\mathlarger{\alpha^{2}_{{\textstyle \text{\tiny QDP}}_{max}}=1}. \label{eq:eq.92}
\end{equation}

At the opposite limit,

\begin{equation}
\mathlarger{ \varepsilon_{{\textstyle \text{\tiny QDP}}_{max}} = \varepsilon_{{\textstyle \text{\tiny QDP}}_c} \bigg|_{\alpha^{2}  \rightarrow 0}. } \label{eq:eq.74}
\end{equation}

From the above analyses, it is concluded that $\varepsilon_{\textstyle \text{\tiny QDP}}$ reaches its maximum at a minimum value of \(\alpha^2\), which is determined by substituting Eqs. \ref{eq:eq.72} and \ref{eq:eq.73} into Eq. \ref{eq:eq.74}, such that

 \begin{align}
    \mathlarger{ \alpha^{2}_{{\textstyle \text{\tiny QDP}}_{min}} = \frac{1}{\theta^{2}}.}
    \label{eq:eq.78}
    \end{align}

\end{itemize}

\subsubsection { Quantum Heating Optimizer}

The transfer of energy from $TR^{h}$ to $TR^{l}$ is a natural process consistent with the second law of thermodynamics and occurs even without the intervention of a QTM. A QHO facilitates this process by prioritizing the release of $E^{l}_{rel}$ to $TR^l$, while absorb $E^{h}_{abs}$ from $TR^h$. Although $E^{out}_{rec}$ from the outside is not the QTM’s primary energy source, its presence enhances the energy released to $TR^l$ without affecting the energy absorbed from $TR^h$, thus improving the QTM overall efficiency.

Similar to QDP, this QTM lacks mention in existing literature. Despite its simpler configuration, understanding all potential QTM designs remains crucial for a comprehensive theoretical investigation.

\begin{itemize}
  \item {QHO Operational Mode} 
    \begin{itemize}
     \item {Energy priority: $E_{target}=E^{l}_{rel}$}.
  
     \item {Energy source: $E_{available}=E^{h}_{abs}$}. 
    \end{itemize}
    
  \item {QHO Efficiency}  
  \begin{align}
    \mathlarger{ \varepsilon_{\textstyle \text{\tiny QHO}} = \frac{\left| E^{l}_{rel}\right|}{\left| E^{h}_{abs}\right|}=-\frac{E^{l}_{rel}}{E^{h}_{abs}} = \frac{1}{\alpha^{2}}.}  \label{eq:eq.75}
    \end{align}

  \item {\textit{QHO Carnot Efficiency} }  
    \begin{equation}
    \begin{aligned}
    \mathlarger{\varepsilon_{{\textstyle \text{\tiny QHO}}_c} =-\frac{E'^{l}_{rel}}{E'^{h}_{abs}} =\frac{T^h}{T^l}=\theta^{2}.}
    \end{aligned}
    \label{eq:eq.63}
    \end{equation}

\item {\textit{Analysis of Limits for $\varepsilon_{\textstyle \text{\tiny QHO}}$ and $\alpha^{2}_{\textstyle \text{\tiny QHO}}$ Values}}

For QHO, $0 < \alpha^{2} < 1$. In Eq. \ref{eq:eq.75} follows that:
    \begin{itemize}
     \item {When $\alpha^{2}  \rightarrow 0$, $\varepsilon_{\textstyle \text{\tiny QHO}} \rightarrow \infty$.}

     \item {When $\alpha^{2}  \rightarrow 1$, $\varepsilon_{\textstyle \text{\tiny QHO}} \rightarrow 1$.}
    \end{itemize}
Therefore, 
\begin{equation}
\mathlarger{\alpha^{2}_{{\textstyle \text{\tiny QHO}}_{max}}=1}, \label{eq:eq.94}
\end{equation}
and

\begin{equation}
\mathlarger{ \varepsilon_{{\textstyle \text{\tiny QHO}}_{max}} = \varepsilon_{{\textstyle \text{\tiny QHO}}_c} \bigg|_{\alpha^{2}  \rightarrow 0}, } \label{eq:eq.33}
\end{equation}

that, from the substituition of Eqs. \ref{eq:eq.72} and \ref{eq:eq.73} into Eq. \ref{eq:eq.33}, gives

 \begin{align}
    \mathlarger{ \alpha^{2}_{{\textstyle \text{\tiny QHO}}_{min}} = \frac{1}{\theta^{2}}.}
    \label{eq:eq.79}
    \end{align}

\end{itemize}

\subsection{Outside Transfers' Operational Region}
This is the most studied and referenced operational region in the literature, as it includes the QEN. However, we propose that it is possible to construct another QTM design in this operational region: the QLL. As can be seen in Subsec. \ref{sec:2LET}-\ref{subsec:Laser}, we believe that QLL more accurately explains how the the laser works.

\subsubsection {Quantum Thermal Engine}

\begin{itemize}
  \item {QEN Operational Mode} 
    \begin{itemize}
     \item {Energy priority: $E_{target}=E^{out}_{ge}$}.
  
     \item {Energy source: $E_{available}=E^{h}_{abs}$}. 
    \end{itemize}
    
  \item {QEN Efficiency}  
\begin{equation}
    \begin{aligned}
    \mathlarger{\varepsilon_{\textstyle \text{\tiny QEN}}} &= \mathlarger{\frac{\left| E^{out}_{ge}\right|}{\left| E^{h}_{abs}\right|}=1 + \frac{E^{l}_{rel}}{E^{h}_{abs}}=1-\frac{1}{\alpha^{2}}.}
    \end{aligned}
    \label{eq:eq.60}
    \end{equation}
    
The efficiency of a monatomic ideal gas as the working medium in a classical thermal engine, $\varepsilon_{CTE}$, operating in the Otto cycle is given by

\begin{equation}
\mathlarger{\varepsilon_{\textstyle \text{\tiny CTE}}=1-\frac{1}{\rho^{\frac{2}{3}}}},
\label{eq:eq.52}
\end{equation}
where $\rho$ is the \textit{compression ratio}, given by

\begin{equation}
\mathlarger{\rho = \frac{\text{\normalsize{volume before compression}}}{\text{\normalsize{volume after compression}}}.}
\label{eq:eq.53}
\end{equation}

By comparing Eqs. \ref{eq:eq.60} and \ref{eq:eq.52}, we observe that $\alpha$ corresponds to the parameter associated with the compression ratio of QEN operating in the Otto cycle. In Sec. \ref{sec:2LET}, when discussing two-level quantum systems as the working medium of QTMs in this cycle, we demonstrate that this conclusion holds for all QTMs.

    \item {\textit{QEN Carnot  Efficiency} }  
        \begin{equation}
          \begin{aligned}
        \mathlarger{\varepsilon_{{\textstyle \text{\tiny QEN}}_c} =1 + \frac{E^{l}_{{rel}_c}}{E^{h}_{{abs}_c}} =1-\frac{T^l}{T^h}= 1-\frac{1}{\theta^{2}}.}
         \end{aligned}
            \label{eq:eq.22}
        \end{equation}

    \item {\textit{Analysis of Limits for $\varepsilon_{\textstyle \text{\tiny QEN}}$ and $\alpha^{2}_{\textstyle \text{\tiny QEN}}$ Values}}
    
    As defined by Eq. \ref{eq:eq.47}, \( \alpha^{2} > 1 \) describes the relationship between \( E^h \) and \( E^l \) for any QTM in the OutTransfers'OR. Based on Eq. \ref{eq:eq.60}, we observe the following behavior:
   
    \begin{itemize}
     \item {When $\alpha^{2}  \rightarrow 1$, $\varepsilon_{\textstyle \text{\tiny QEN}} \rightarrow 0$.}

     \item {When $\alpha^{2}  \rightarrow \infty$, $\varepsilon_{\textstyle \text{\tiny QEN}} \rightarrow 1$.}
    \end{itemize}
    
This leads us to conclude that, since $\varepsilon_{\textstyle \text{\tiny QEN}} \rightarrow 0$ is allowed, 

\begin{equation}
\mathlarger{\alpha^{2}_{{\textstyle \text{\tiny QEN}}_{min}}=1}. \label{eq:eq.93}
\end{equation}

However, as is well known,  $\varepsilon_{\textstyle \text{\tiny QEN}} \rightarrow 1$ implies  $E^{l}_{{rel}} \rightarrow 0$, which contradicts the second law of thermodynamics for a thermal machine. The efficiency limit is determined by the ideal capacity of the TRs to supply and absorb energy from the working medium. Therefore, 

\begin{equation}
\mathlarger{ \varepsilon_{{\textstyle \text{\tiny QEN}}_{max}} = \varepsilon_{{\textstyle \text{\tiny QEN}}_c} \bigg|_{\alpha^{2}  \rightarrow \infty}, } \label{eq:eq.62}
\end{equation}
and

\begin{align}
    \mathlarger{ \alpha^{2}_{{\textstyle \text{\tiny QEN}}_{max}} = \theta^{2}.}
    \label{eq:eq.80}
    \end{align}
    
\end{itemize}

\subsubsection {Quantum Thermal Laser-Like}
\begin{itemize}
  \item {QLL Operational Mode} 
    \begin{itemize}
     \item {Energy priority: $E_{target}=E^{l}_{rel}$}.
  
     \item {Energy source: $E_{available}=E^{h}_{abs}$}. 
    \end{itemize}
    
  \item {QLL Efficiency}  
    \begin{equation}
    \mathlarger{\varepsilon_{\textstyle \text{\tiny {QLL}}}=
    \frac{\left|E^{l}_{rel}\right|}{\left|
    E^{h}_{abs}\right|}=- \frac{E^{l}_{rel}}{E^{h}_{abs}}=\frac{1}{\alpha^{2}}.}
    \label{eq:eq.04}
    \end{equation}
    
  \item {\textit{QLL Carnot Efficiency} }  
 \begin{equation}
\mathlarger{\varepsilon_{{\textstyle \text{\tiny {QLL}}}_c}=- \frac{E^{l}_{{rel}_c}}{E^{h}_{{abs}_c}}= \frac{T^l}{T^h}=\frac{1}{\theta^{2}}.}
    \label{eq:eq.06}
    \end{equation}

\item {\textit{Analysis of Limits for $\varepsilon_{\textstyle \text{\tiny QLL}}$ and $\alpha^{2}_{\textstyle \text{\tiny QLL}}$ Values}}

For QLL, $\alpha^{2} > 1$. In Eq. \ref{eq:eq.04} follows that:
    \begin{itemize}
     \item {When $\alpha^{2}  \rightarrow 1$, $\varepsilon_{\textstyle \text{\tiny QLL}} \rightarrow 1$.}

     \item {When $\alpha^{2}  \rightarrow \infty$, $\varepsilon_{\textstyle \text{\tiny QLL}} \rightarrow 0$.}
    \end{itemize}

QLL is the only QTM for which $\varepsilon_{{\textstyle \text{\tiny QLL}}_c}$ must bound the minimum value of \(\varepsilon_{\textstyle \text{\tiny QLL}}\), as $E^{l}_{{rel}} \rightarrow 0$, when $\varepsilon_{\textstyle \text{\tiny QLL}} \rightarrow 0$.

Therefore, 

\begin{equation}
\mathlarger{ \varepsilon_{{\textstyle \text{\tiny QLL}}_{min}} = \varepsilon_{{\textstyle \text{\tiny QLL}}_c} \bigg|_{\alpha^{2}  \rightarrow \infty}. } \label{eq:eq.10}
\end{equation}

The data above indicate that $\varepsilon_{\textstyle \text{\tiny QLL}}$ reaches its minimum at the maximum value of \(\alpha^2\). This value can be determined by substituting Eqs. \ref{eq:eq.04} and \ref{eq:eq.06} into Eq. \ref{eq:eq.10}, as follows

 \begin{align}
    \mathlarger{ \alpha^{2}_{{\textstyle \text{\tiny QLL}}_{max}} = \theta^{2}.}
    \label{eq:eq.81}
    \end{align}
    
Conversely, there are no constraints on $\varepsilon_{\textstyle \text{\tiny QLL}} \rightarrow 1$, and thus,

\begin{equation}
\mathlarger{\alpha^{2}_{{\textstyle \text{\tiny QLL}}_{min}}=1}. \label{eq:eq.95}
\end{equation}

\end{itemize}

\subsection{Thermal Pumpers' Operational Region}
As previously mentioned, in the Pumpers'OR, energy transfer between the TRs does not follow the thermodynamically natural path. The presence of $E^{out}_{rec}$ is crucial, as it enables the QTM to extract energy from $TR^l$ and release it into $TR^h$. The QRE and QHP are the QTMs operating within the Pumpers'OR. While QRE is well established in the current literature, similar to QCO and QHT, a proper definition of QHP’s characteristics and the use of standardized nomenclature are still required. 

\subsubsection {Quantum Refrigerator}

\begin{itemize}
  \item {QRE Operational Mode} 
    \begin{itemize}
     \item {Energy priority: $E_{target}=E^{l}_{abs}$}.
  
     \item {Energy source: $E_{available}=E^{out}_{rec}$}. 
    \end{itemize}
    
  \item {QRE Efficiency} 
    \begin{equation}
    \begin{aligned}
    \mathlarger{\varepsilon_{\textstyle \text{\tiny QRE}}} &= \mathlarger{\frac{\left| E^{l}_{abs}\right|}{\left| E^{out}_{rec}\right|}=-\frac{1}{1+\frac{E^{h}_{rel}}{E^{l}_{abs}}} = \frac{1}{\alpha^{2}-1}.}
    \end{aligned}
    \label{eq:eq.61}
    \end{equation}

  \item {\textit{QRE Carnot Efficiency} }  
    \begin{equation}
\mathlarger{\varepsilon_{{\textstyle \text{\tiny QRE}}_c}=-\frac{1}{1+\frac{E^{h}_{{rel}_c}}{E^{l}_{{abs}_c}}}=\frac{1}{\frac{T^h}{T^l}-1}=\frac{1}{\theta^{2}-1}.} \label{eq:eq.59}
 \end{equation} 	  

\item {\textit{Analysis of Limits for $\varepsilon_{\textstyle \text{\tiny QRE}}$ and $\alpha^{2}_{\textstyle \text{\tiny QRE}}$ Values}}

For QRE, $\alpha^{2} > 1$. In Eq. \ref{eq:eq.58} follows that:
    \begin{itemize}
     \item {When $\alpha^{2}  \rightarrow 1$, $\varepsilon_{\textstyle \text{\tiny QRE}} \rightarrow \infty$.}

     \item {When $\alpha^{2}  \rightarrow \infty$, $\varepsilon_{\textstyle \text{\tiny QRE}} \rightarrow 0$.}
    \end{itemize}

In this case, we have 

\begin{equation}
\mathlarger{\alpha^{2}_{{\textstyle \text{\tiny QRE}}_{max}}\rightarrow \infty}, \label{eq:eq.96}
\end{equation}
 
as there are no constraints on $\varepsilon_{\textstyle \text{\tiny QRE}} \rightarrow 0$.
    
At the opposite limit, we have 

\begin{equation}
\mathlarger{ \varepsilon_{{\textstyle \text{\tiny QRE}}_{max}} = \varepsilon_{{\textstyle \text{\tiny QRE}}_c} \bigg|_{\alpha^{2}  \rightarrow 1}. } \label{eq:eq.49}
\end{equation}

As is obtained for QTMs in the 2Acq$^{h}$ subregion, $\varepsilon_{\textstyle \text{\tiny QRE}}$ reaches its maximun at a minimum value of \(\alpha^2\), substituting Eqs. \ref{eq:eq.61} and \ref{eq:eq.59} into Eq. \ref{eq:eq.49}, such that

 \begin{align}
    \mathlarger{ \alpha^{2}_{{\textstyle \text{\tiny QRE}}_{min}} = \theta^{2}.}
    \label{eq:eq.82}
    \end{align}

\end{itemize}

\subsubsection {Quantum Heat Pumper}

\begin{itemize}
  \item {QHP Operational Mode} 
    \begin{itemize}
     \item {Energy priority: $E_{target}=E^{h}_{rel}$}.
  
     \item {Energy source: $E_{available}=E^{out}_{rec}$}. 
    \end{itemize}
    
  \item {QHP Efficiency}  
  \begin{align}
   \mathlarger{ \varepsilon_{\textstyle \text{\tiny QHP}} = \frac{\left| E^{h}_{rel}\right|}{\left| E^{out}_{rec}\right|}=\frac{1}{1+\frac{E^{l}_{abs}}{E^{h}_{rel}}} = \frac{1}{1-\frac{1}{\alpha^{2}}}.} \label{eq:eq.58}
    \end{align}
    
  \item {\textit{QHP Carnot Efficiency} }  
    \begin{equation} \mathlarger{\varepsilon_{{\textstyle \text{\tiny QHP}}_c} =\frac{1}{1+\frac{E^{l}_{{abs}_c}}{E^{h}_{{rel}_c}}}=\frac{1}{1-\frac{T^l}{T^h}}=\frac{1}{1-\frac{1}{\theta^{2}}}.} \label{eq:eq.64}
    \end{equation} 	  

\item {\textit{Analysis of Limits for $\varepsilon_{\textstyle \text{\tiny QHP}}$ and $\alpha^{2}_{\textstyle \text{\tiny QHP}}$ Values}}

For QHP, $\alpha^{2} > 1$. In Eq. \ref{eq:eq.58} follows that:
    \begin{itemize}
     \item {When $\alpha^{2}  \rightarrow 1$, $\varepsilon_{\textstyle \text{\tiny QHP}} \rightarrow \infty$.}

     \item {When $\alpha^{2}  \rightarrow \infty$, $\varepsilon_{\textstyle \text{\tiny QHP}} \rightarrow 1$.}
    \end{itemize}
    
Therefore, as the previous analyses, we have

\begin{equation}
\mathlarger{\alpha^{2}_{{\textstyle \text{\tiny QHP}}_{max}}\rightarrow \infty}, \label{eq:eq.97}
\end{equation}

\begin{equation}
\mathlarger{ \varepsilon_{{\textstyle \text{\tiny QHP}}_{max}} = \varepsilon_{{\textstyle \text{\tiny QHP}}_c} \bigg|_{\alpha^{2}  \rightarrow 1}, } \label{eq:eq.51}
\end{equation}
and

 \begin{align}
    \mathlarger{ \alpha^{2}_{{\textstyle \text{\tiny QHP}}_{min}} = \theta^{2}.}
    \label{eq:eq.83}
    \end{align}

\end{itemize}

\subsection{Efficiency Relationships Within the Same Operational Region}

The relationship between the efficiencies of QTMs in each operational region can be easily inferred from the schematic representation of the QTMs in Fig. \ref{fig:esquemaQTM}(b to d). Furthermore, the expressions derived for the efficiencies of all QTMs allow for accurate deductions of these relationships, as described below.

\begin{itemize}

       \item {\textit{Bi-Acquirers' Operational Region}}

    From Eqs. \ref{eq:eq.67} and \ref{eq:eq.69}, and from Eqs. \ref{eq:eq.72} and \ref{eq:eq.75}, we obtain
    
    \begin{equation} 
    \mathlarger{\varepsilon_{\textstyle \text{\tiny QHT}} - \varepsilon_{\textstyle \text{\tiny QCO}} = \frac{1}{1-\alpha^{2}} - \frac{1}{\frac{1}{\alpha^{2}}-1} = 1},  \label{eq:eq.25}
    \end{equation} 
    and
    
      \begin{equation} \mathlarger{\varepsilon_{\textstyle \text{\tiny QHO}} - \varepsilon_{\textstyle \text{\tiny QDP}} = \frac{1}{\alpha^{2}} - \left( \frac{1}{\alpha^{2}}-1\right)= 1},  \label{eq:eq.86}
    \end{equation} 
which leads to

    \begin{equation} 
    \mathlarger{\varepsilon_{\textstyle \text{\tiny QHT}} > \varepsilon_{\textstyle \text{\tiny QCO}}},      \label{eq:eq.36}
    \end{equation}
and

    \begin{equation} 
    \mathlarger{\varepsilon_{\textstyle \text{\tiny QHO}} > \varepsilon_{\textstyle \text{\tiny QDP}}.}      \label{eq:eq.88}
    \end{equation}

    \item {\textit{Outside Transfers' Operational Region}}

    The efficiencies of QTMs in the OutTransfers'OR have a different relationship. From Eqs. \ref{eq:eq.60} and \ref{eq:eq.04}, we have
    
    \begin{equation} 
    \mathlarger{\varepsilon_{\textstyle \text{\tiny QEN}} + \varepsilon_{\textstyle \text{\tiny QLL}} = 1 - \frac{1}{\alpha^{2}} + \frac{1}{\alpha^{2}} = 1.}  \label{eq:eq.27}
    \end{equation} 

    In this case, there is no dominance of one $\varepsilon_{\textstyle \text{\tiny QTM}}$ over the other. However, Eq. \ref{eq:eq.27} implies that neither QEN nor QLL can have an efficiency greater than one.

    \item {\textit{Thermal Pumpers' Operational Region}}
    
    From Eqs. \ref{eq:eq.61} and \ref{eq:eq.58}, we get
    
    \begin{equation} 
    \mathlarger{\varepsilon_{\textstyle \text{\tiny QHP}} - \varepsilon_{\textstyle \text{\tiny QRE}} = \frac{1}{1-\frac{1}{\alpha^{2}}} - \frac{1}{\alpha^{2}-1} = 1},  \label{eq:eq.26}
    \end{equation}

leading to

    \begin{equation} 
    \mathlarger{\varepsilon_{\textstyle \text{\tiny QHP}} > \varepsilon_{\textstyle \text{\tiny QRE}}.}  \label{eq:eq.28}
    \end{equation} 

\subsection{Thermal High-Low Energy Ratio Values at Operational Region Intersections}

   Based on the results obtained in the previous subsections for the limit $\alpha^{2}$ values of each QTM, it is possible to define the values that determine the intersections between two operational regions, including the intersection between the two subregions within the 2Acquirers'OR.
    
\begin{itemize}

       \item {\textit{$2Acq^{out} - 2Acq^{h}$ Intersection}} 

By comparing Eqs. \ref{eq:eq.76}, \ref{eq:eq.77}, \ref{eq:eq.78}, and \ref{eq:eq.79}, we conclude that the $\alpha^{2}$ value in $2Acq^{out} - 2Acq^{h}$ subregions intersection has to be

   \begin{equation} 
    \mathlarger{\alpha^{2}_{2Acq^{out} \cap 2Acq^{h}} = \frac{1}{\theta^{2}}.}  \label{eq:eq.84}
    \end{equation} 

 \item {\textit{$2Acquirers - OutTransfers$ Intersection}} 

The $\alpha^{2}$ value in $2Acquirers - OutTransfers$ operational regions intersection is 
 \begin{equation} 
    \mathlarger{\alpha^{2}_{2Acq \cap OutT} = 1,}  \label{eq:eq.85}
    \end{equation} 
since $\alpha^{2}_{{\textstyle \text{\tiny 2Acq}}_{max}} =\alpha^{2}_{{\textstyle \text{\tiny OutT}}_{min}}=1$ (see Eqs. \ref{eq:eq.92}, \ref{eq:eq.94}, \ref{eq:eq.93}, and \ref{eq:eq.95}).
    
 \item {\textit{$OutTransfers - Pumpers$ Intersection}} 
 
From Eqs. \ref{eq:eq.80}, \ref{eq:eq.81}, \ref{eq:eq.82} and \ref{eq:eq.83}, we have

   \begin{equation} 
    \mathlarger{\alpha^{2}_{OutT \cap Pump}= \theta^{2}.} \label{eq:eq.87}
    \end{equation} 
    
    \end{itemize}

\end{itemize}

\begin{table*}[ht]
\centering
\large
\caption{Summary of operational regions and their corresponding potential QTMs: essential information for any QTM using a working medium operating between \(TR^h\) and \(TR^l\), governed by principles analogous to those applied in this work.}
    \includegraphics[width=1.0\textwidth]{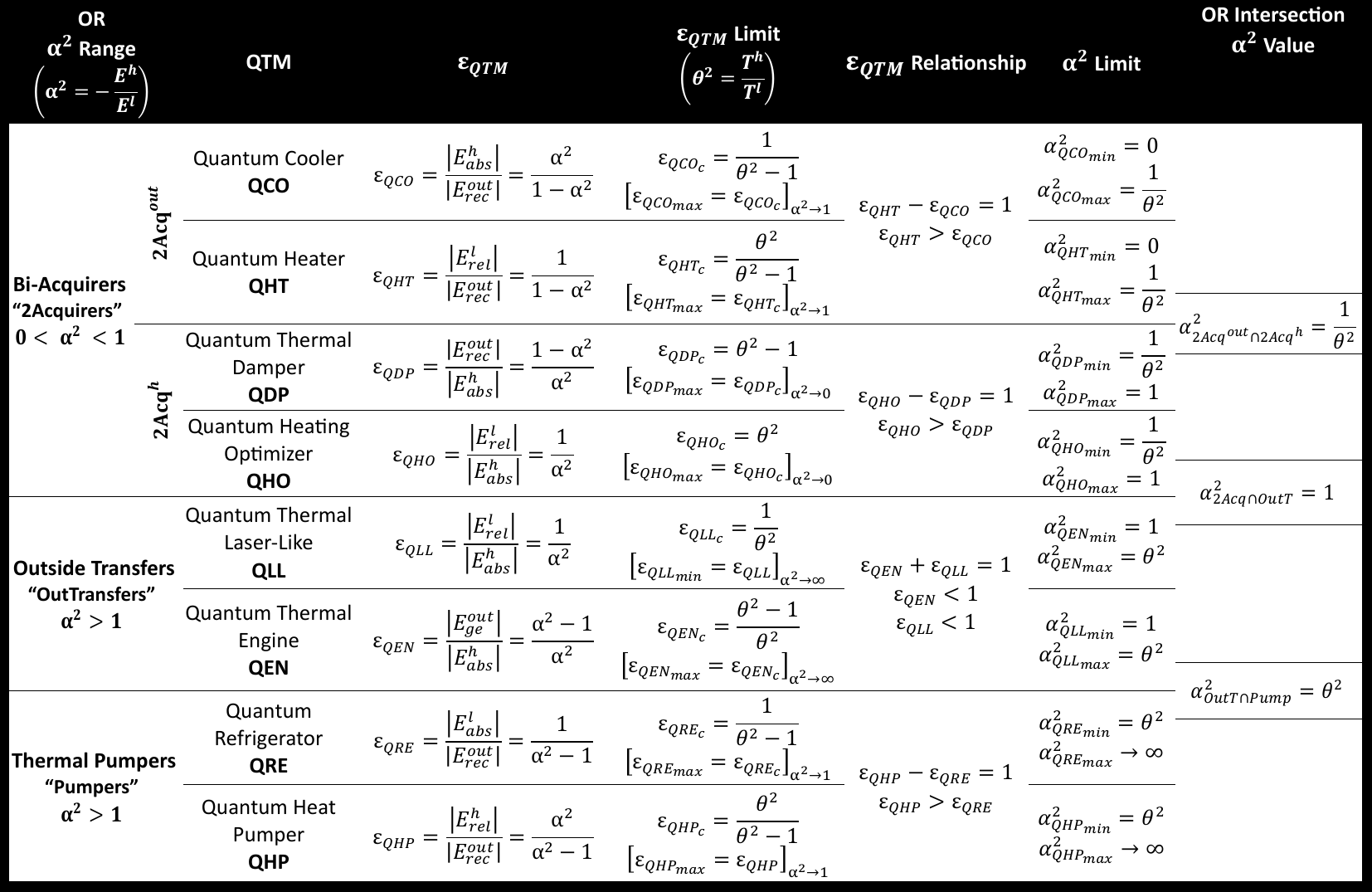} 
    \label{tab:all_QTMs_table}
\end{table*}

The Table \ref{tab:all_QTMs_table} summarizes the information derived from our study, which we consider applicable to any working medium operating between \(TR^h\) and \(TR^l\), governed by principles similar to those utilized in this work.

In the next section, we use these results to examine the two-level quantum systems as the working medium of QTMs operating in the Otto cycle.

 \section{Two-Level One-Particle Quantum System}
 \label{sec:2LET}

\begin{figure*}[ht]
    \centering
    \includegraphics[scale=0.32]{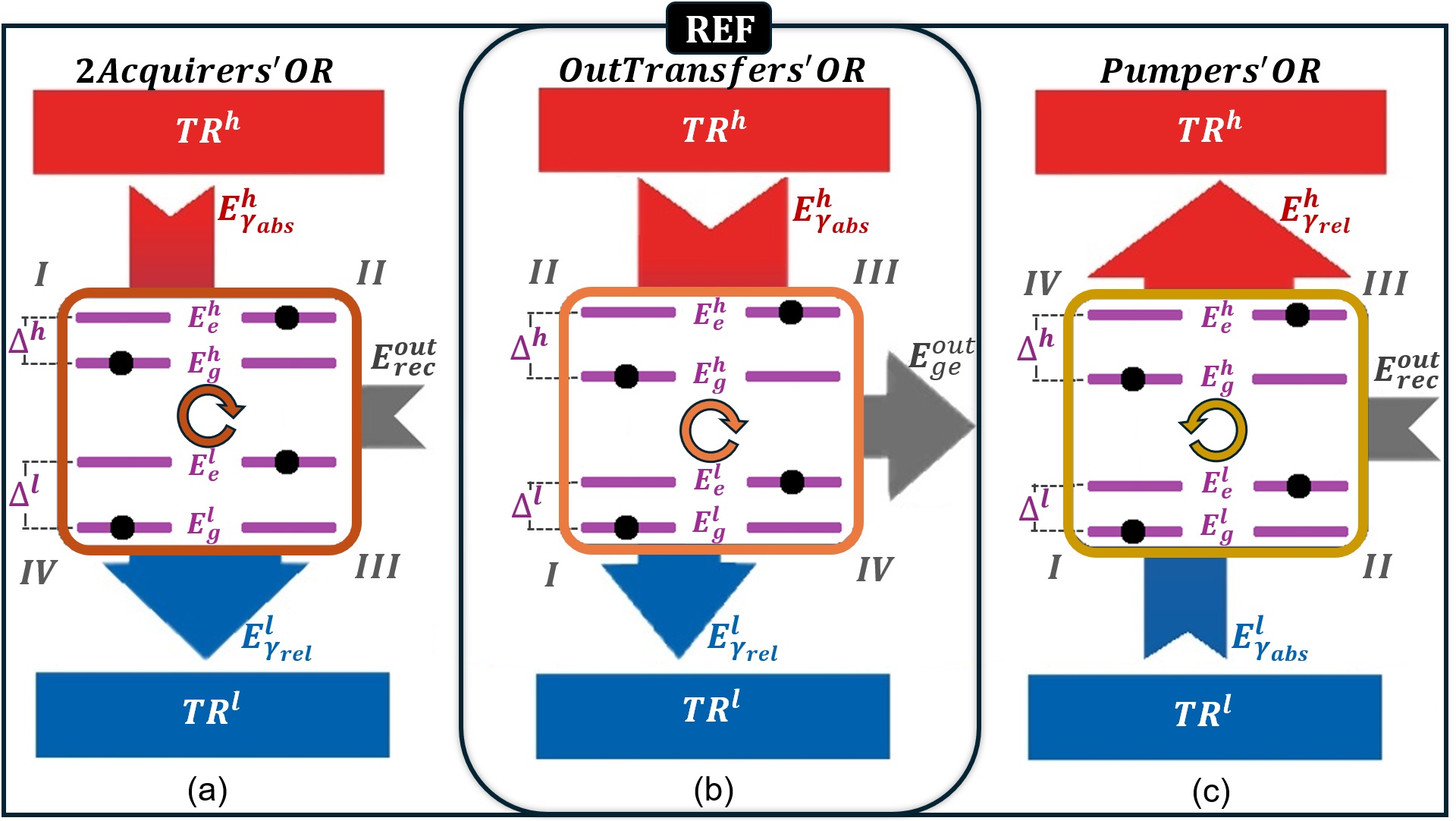}
    \caption{Configurations of a two-level quantum system used as a working medium in its three distinct operational regions: (a) 2Acquirers'OR, (b) OutTransfers'OR, and (c) Pumpers'OR. For each configuration, there are four strokes, transitioning the system through four distinct states. $E^l_g$ and $E^l_e$ represent the ground and excited energy eigenvalues of the two-level quantum system, respectively, with an energy gap of $\Delta^{l}$, while $E^h_g$ and $E^h_e$ denote the ground and excited energy eigenvalues for an energy gap of $\Delta^{h}$. The strokes involve energy exchanges with the TRs, either absorbing or releasing $E^{h}_{\gamma}$ to/from $TR^{h}$ and absorbing or releasing $E^{l}_{\gamma}$ to/from $TR^{l}$, as well as involving changes in the dimensional configurations of the two-level quantum system, during which it either receives or generates $E^{out}$.}
    \label{fig:esquema2LET}
\end{figure*}

Based on the proposed operational regions and subregions, Fig. \ref{fig:esquema2LET} illustrates the possible configurations of a two-level quantum system as the working medium of a QTM. Figures \ref{fig:esquema2LET}a, \ref{fig:esquema2LET}b, and \ref{fig:esquema2LET}c display the 2Acquirers'OR, OutTransfers'OR, and Pumpers'OR configurations of the two-level quantum system, respectively.  

We treat the OutTransfers'OR configuration as the reference (Fig. \ref{fig:esquema2LET}b), from which subsequent calculations naturally evolve.

As illustrated in Fig. \ref{fig:esquema2LET}b, the two-level quantum system has four possible states. In state I, the particle is in the ground state with energy $E^l_g$, while the energy gap between the levels, $\Delta^{l}$, is given by

    \begin{equation}
    \mathlarger{\Delta^{l}= E^l_e-E^l_g}, \label{eq:eq.11}
    \end{equation}
where \( E^l_e \) is the excited energy level of the system in state I. In state II, the particle occupies the ground energy level of a different dimensional configuration, with ground and excited energy levels denoted as \( E^h_g \) and \( E^h_e \), respectively. These levels are separated by an energy gap \( \Delta^h \), defined as

    \begin{equation}
    \mathlarger{\Delta^{h}= E^h_e-E^h_g}. \label{eq:eq.16}
    \end{equation}

 In state III, the system retains the dimensional configuration of state II, but the particle is now in the excited state, $E^h_e$. The system returns to the dimensional configuration of state I, with the particle occupying the excited state $E^l_e$, in state IV.

As detailed in Subsec. \ref{sec:2LET}-\ref{subsec: Otto Cycle}, changes in the dimensional configuration of the two-level quantum system are driven by energy exchange with the QTM's outside ($E^{out}$), while transitions of the particle between energy levels occur through the absorption or emission of photons. These transitions correspond to energy exchanges with $TR^h$, denoted as $E^{h}_{\gamma}$, and with $TR^l$, denoted as $E^{l}_{\gamma}$.

In its OutTranfers'OR, the system reaches the maximum temperature $T_{max} = T^{h}$ after absorbing energy $E^{h}_{\gamma}$ from $TR^h$, i.e., when it is in state III. Conversely, it reaches the minimum temperature $T_{min} = T^{l}$ after releasing energy $E^{l}_{\gamma}$ to $TR^l$, in state I.

The state labels in the three operational regions were defined so that state III (I) corresponds to the configuration with the largest (smallest) energy level separation, with the particle in the excited (ground) state at temperature \(T^h\) (\(T^l\)) \footnote{It is important to note that, as analyzed in Subsec. \ref{subsec: Carnot Cycle}, in its 2Acquirers'OR, the system reaches its maximum ($T^h$) or minimum ($T^{l}$) temperature immediately before releasing ($T^h$ to $TR^{l}$) or absorbing ($T^{l}$ from $TR^{h}$) energy, corresponding to state III ($T^h$) or state I ($T^{l}$) in Fig. \ref{fig:esquema2LET}a.}.
 
\subsection{Otto Cycle}
\label{subsec: Otto Cycle}

A classical thermal machine operating in the Otto cycle goes through a cycle composed of two adiabatic and two isochoric strokes. During the adiabatic strokes, no heat is exchanged with the surroundings, and energy changes occur only due to volume variations. In the isochoric strokes, the working medium exchanges heat with $TR^h$ and $TR^l$, while maintaining a constant volume.

One of the distinguishing features of this cycle is that each stroke involves either work or heat exchange, but never both simultaneously. This is particularly valuable for the theorical study of QTMs, as each stroke involves either an \(E^{TR}\) or an \(E^{out}\) exchange, making it easier to achieve satisfactory results.

In any thermodynamic cycle of a QTM, a wide range of outcomes is possible, largely influenced by the choice of the working medium, which in this case is a two-level quantum system. For this specific working medium, these outcomes are determined by solving the Schrödinger equation to identify the energy levels associated with quantum particle transitions during the QTM cycle.

The multilevel quantum systems as the working medium of the QEN operating in the Otto cycle, was carefully analyzed by Quan et al. \cite{Quan2007}. They identified the quantum analogs of heat exchange, $\delta{Q}$, and work performed, $\delta{W}$, for a multilevel working medium with energy eigenvalues $E_n$, where $n=1,2,3,...$. We applied their findings to our study of QTMs, utilizing the implications of Eqs. \ref{eq:eq.100} and \ref{eq:eq.101}, and the labels from Fig. \ref{fig:esquema2LET}b, as follows

    \begin{equation}
     \mathlarger{\delta{Q_{abs}}=\delta{E^{TR}_{\gamma_{abs}}}=\left[\sum_n E_n d\mathcal{P}_n\right]^{TR}_{\gamma_{abs}}},  \label{eq:eq.03}
    \end{equation}
    
    \begin{equation}
     \mathlarger{\delta{Q_{rel}}=\delta{E^{TR}_{\gamma_{rel}}}=\left[\sum_n E_n d\mathcal{P}_n\right]^{TR}_{\gamma_{rel}}}, \label{eq:eq.89}
    \end{equation}
    
and

    \begin{equation}
     \mathlarger{\delta{W}=\delta{E^{out}}=\sum_n  \mathcal{P}_n dE_n.} \label{eq:eq.05}
    \end{equation}
where, in our quantum systems, \( n=g \) and \( n=e \) denote the ground and excited energy levels, respectively.

$\mathcal{P}_n$ is the occupation probability of the $n$-th energy eigenstate, given by

    \begin{equation}
     \mathlarger{\mathcal{P}_{n} = \frac{e^{-\frac{E_{n}}{k_B T}}}{Z}}, \label{eq:eq.15}
    \end{equation}
where
    \begin{equation}
      \mathlarger{Z = \sum_n e^{-\frac{E_{n}}{k_B T}}} \label{eq:eq.31}
    \end{equation} 
is the system partition function at temperature $T$.

 Considering that the strokes are quasi-static processes, from Eq. \ref{eq:eq.03}, we can obtain the energy $E^{h}_{\gamma_{abs}}$ for the OutTransfers'OR in Fig. \ref{fig:esquema2LET}b, as

    \begin{equation}
     \begin{split}
        \mathlarger{E^{h}_{\gamma_{abs}}}& \mathlarger{= \sum_n \int_{II}^{III} E_n d\mathcal{P}_n=\sum_{n=e,g} E^h_n\left[\mathcal{P}_n^{(III)} - \mathcal{P}_n^{(II)}\right]} \\
       & \mathlarger{= E^h_e\left[\mathcal{P}_e^{(III)} - \mathcal{P}_e^{(II)}\right]+E^h_g\left[\mathcal{P}_g^{(III)} - \mathcal{P}_g^{(II)}\right].}
    \end{split}
    \label{eq:eq.12}
   \end{equation}

In turn, starting from Eq. \ref{eq:eq.03}, the expression for $E^{l}_{\gamma_{rel}}$ is given by

    \begin{equation}
    \begin{split}
        \mathlarger{E^{l}_{\gamma_{rel}}} & \mathlarger{= \sum_n \int_{IV}^{I} E_n d\mathcal{P}_n=\sum_{n=e,g} E^l_n\left[\mathcal{P}_n^{(IV)} - \mathcal{P}_n^{(III)}\right]} \\
        & \mathlarger{= E^l_e\left[\mathcal{P}_e^{(I)} - \mathcal{P}_e^{(IV)}\right]+E^l_g\left[\mathcal{P}_g^{(I)} - \mathcal{P}_g^{(IV)}\right].}
    \end{split}
        \label{eq:eq.19}
    \end{equation}

Since this is an Otto cycle, strokes I-II and III-IV are carried out without energy exchange with TR$^h$ or TR$^l$, so, for the occupation probability, we have 

    \begin{equation}
    \mathlarger{\mathcal{P}_n^{(II)} = \mathcal{P}_n^{(I)}=\mathcal{P}^l_n}, \label{eq:eq.13}
    \end{equation}
and

    \begin{equation}
    \mathlarger{\mathcal{P}_n^{(IV)} = \mathcal{P}_n^{(III)}=\mathcal{P}^h_n.} \label{eq:eq.14}
    \end{equation}

The choice of $\mathcal{P}^l_n$ and $\mathcal{P}^h_n$ implies that the calculation will be performed using the information from states I and III, respectively, since the temperatures are know, with $T^{(I)}=T^{l}$ and $T^{(III)}=T^{h}$.

Thus, we obtain the expressions

    \begin{equation}
    \begin{split}
        \mathlarger{E^{h}_{\gamma_{abs}}}& \mathlarger{= E^h_e\left[\mathcal{P}^h_e - \mathcal{P}^l_e\right]+E^h_g\left[\mathcal{P}^h_g - \mathcal{P}^l_g\right]=}  \\
        & \mathlarger{=\Delta^{h}\frac{\left[e^{-\frac{E^l_g}{k_{\textstyle \text{\tiny B}}T^{l}}} e^{-\frac{E^h_e}{k_{\textstyle \text{\tiny B}}T^{h}}}   -  e^{-\frac{E^h_g}{k_{\textstyle \text{\tiny B}}T^{h}}} e^{-\frac{E^l_e}{k_{\textstyle \text{\tiny B}}T^{l}}}\right]}{\left[e^{-\frac{E^h_g}{k_{\textstyle \text{\tiny B}}T^{h}}}+e^{-\frac{E^h_e}{k_{\textstyle \text{\tiny B}}T^{h}}}\right] \left[e^{-\frac{E^l_g}{k_{\textstyle \text{\tiny B}}T^{l}}}+e^{-\frac{E^l_e}{k_{\textstyle \text{\tiny B}}T^{l}}}\right]}},
    \end{split}
    \label{eq:eq.20}
    \end{equation}
and

    \begin{equation}
    \begin{split}
        \mathlarger{E^{l}_{\gamma_{rel}}} & \mathlarger{= E^l_e\left[\mathcal{P}^l_e - \mathcal{P}^h_e\right]+E^l_g\left[\mathcal{P}^l_g - \mathcal{P}^h_g\right]=} \\
        & \mathlarger{= -\Delta^{l}\frac{\left[e^{-\frac{E^l_g}{k_{\textstyle \text{\tiny B}}T^{l}}} e^{-\frac{E^h_e}{k_{\textstyle \text{\tiny B}}T^{h}}}   -  e^{-\frac{E^h_g}{k_{\textstyle \text{\tiny B}}T^{h}}} e^{-\frac{E^l_e}{k_{\textstyle \text{\tiny B}}T^{l}}}\right]}{\left[e^{-\frac{E^h_g}{k_{\textstyle \text{\tiny B}}T^{h}}}+e^{-\frac{E^h_e}{k_{\textstyle \text{\tiny B}}T^{h}}}\right] \left[e^{-\frac{E^l_g}{k_{\textstyle \text{\tiny B}}T^{l}}}+e^{-\frac{E^l_e}{k_{\textstyle \text{\tiny B}}T^{l}}}\right]},}
    \end{split}
    \label{eq:eq.23}
    \end{equation}
where $\Delta^{l}$ and $\Delta^{h}$ are the energy gaps between the levels defined in Eq. \ref{eq:eq.11} and Eq. \ref{eq:eq.16}, respectively. The Eq. \ref{eq:eq.23} yields negative values, which is expected since we define the energies released to the TRs as negative. 

For QTMs, it is usually simpler to first solve Eqs. \ref{eq:eq.20} and \ref{eq:eq.23}, and then to determine \(E^{out}_{ge}\). Thus, using Eq. \ref{eq:eq.18}, we have

    \begin{equation}
     \mathlarger{E^{out}_{ge}= E^{h}_{\gamma_{abs}} +  E^{l}_{\gamma_{rel}}}. \label{eq:eq.29}
    \end{equation}

The calculations from Eq. \ref{eq:eq.12} to Eq. \ref{eq:eq.29} can be similarly applied to the QTMs in the 2Acquirers'OR (Fig. \ref{fig:esquema2LET}a), and Pumpers'OR (Fig. \ref{fig:esquema2LET}c), yielding equations analogous to Eqs. \ref{eq:eq.20} to \ref{eq:eq.29}. Thus, we can generalize these results as follows:

    \begin{equation}
    \mathlarger{E^{h}_{\gamma}=\Delta^{h}\frac{\left[e^{-\frac{1}{k_{\textstyle \text{\tiny B}}T^{l}}E^l_g} e^{-\frac{1}{k_{\textstyle \text{\tiny B}}T^{l}} \frac{E^h_e}{\theta^2}}   -  e^{-\frac{1}{k_{\textstyle \text{\tiny B}}T^{l}} \frac{E^h_g}{\theta^2}} e^{-\frac{1}{k_{\textstyle \text{\tiny B}}T^{l}}E^l_e}\right]}{\left[e^{-\frac{1}{k_{\textstyle \text{\tiny B}}T^{l}} \frac{E^h_g}{\theta^2}}+e^{-\frac{1}{k_{\textstyle \text{\tiny B}}T^{l}} \frac{E^h_e}{\theta^2}}\right] \left[e^{-\frac{1}{k_{\textstyle \text{\tiny B}}T^{l}}E^l_g}+e^{-\frac{1}{k_{\textstyle \text{\tiny B}}T^{l}}E^l_e}\right]},}  \label{eq:eq.45}
    \end{equation}

    \begin{equation}
    \mathlarger{E^{l}_{\gamma}=-\Delta^{l}\frac{\left[e^{-\frac{1}{k_{\textstyle \text{\tiny B}}T^{l}}E^l_g} e^{-\frac{1}{k_{\textstyle \text{\tiny B}}T^{l}} \frac{E^h_e}{\theta^2}}   -  e^{-\frac{1}{k_{\textstyle \text{\tiny B}}T^{l}} \frac{E^h_g}{\theta^2}} e^{-\frac{1}{k_{\textstyle \text{\tiny B}}T^{l}}E^l_e}\right]}{\left[e^{-\frac{1}{k_{\textstyle \text{\tiny B}}T^{l}} \frac{E^h_g}{\theta^2}}+e^{-\frac{1}{k_{\textstyle \text{\tiny B}}T^{l}} \frac{E^h_e}{\theta^2}}\right] \left[e^{-\frac{1}{k_{\textstyle \text{\tiny B}}T^{l}}E^l_g}+e^{-\frac{1}{k_{\textstyle \text{\tiny B}}T^{l}}E^l_e}\right]}}  \label{eq:eq.46}
    \end{equation}
and

    \begin{equation}
     \mathlarger{E^{out}= E^{h} +  E^{l}}, \label{eq:eq.30}
    \end{equation}
where, in Eqs. \ref{eq:eq.45} and \ref{eq:eq.46}, we used Eq. \ref{eq:eq.24} to substitute \(T^h\) with \(\theta^2 T^l\). Additionally, as expected, Eq. \ref{eq:eq.30} aligns with Eq. \ref{eq:eq.18}, since the latter already provides the generic expression for the relationship between energies in any operational region. Positive (absorption or generation) and negative (release or receipt) energy values naturally emerge from the relationship between them in each operational region of the working medium.

The ratio between Eqs. \ref{eq:eq.45} and \ref{eq:eq.46} is given by

    \begin{equation}
    \mathlarger{\frac{E^{h}_{\gamma}}{E^{l}_{\gamma}} = -\frac{\Delta^{h}}{\Delta^{l}},} \label{eq:eq.41}
    \end{equation}
which, compared to Eq. \ref{eq:eq.47}, gives

    \begin{equation}
    \mathlarger{\frac{\Delta^{h}}{\Delta^{l}}=\alpha^{2}.}   \label{eq:eq.44}
    \end{equation}

The Eq. \ref{eq:eq.44} shows that $\alpha^{2}$, as expected, is the parameter related to the compression ratio $\rho$ (see Eq. \ref{eq:eq.53}) for any two-level quantum system as the working medium of QTMs operating in the Otto cycle. Furthermore, by applying Eqs. \ref{eq:eq.11}, \ref{eq:eq.16}, and \ref{eq:eq.45} to \ref{eq:eq.44}, the designs of all QTMs proposed in this work, along with their characteristics summarized in Table \ref{tab:all_QTMs_table}, remain valid, as long as the energy eigenvalues of the two-level quantum system are known.

\subsection{Rethinking the Laser: Beyond the Quantum Engine Classification}
  \label{subsec:Laser}
More than sixty years ago, Scovil and Schultz-DuBois \cite{Scovil1959} proposed that the operation of a maser (microwave amplification by stimulated emission of radiation) could be evaluated similarly to a classical thermal engine. The laser (light amplification by stimulated emission of radiation), which directly succeed the maser, should also operate on the same fundamental principle. 

\begin{figure}[ht]
        \centering
        \includegraphics[scale=0.32]{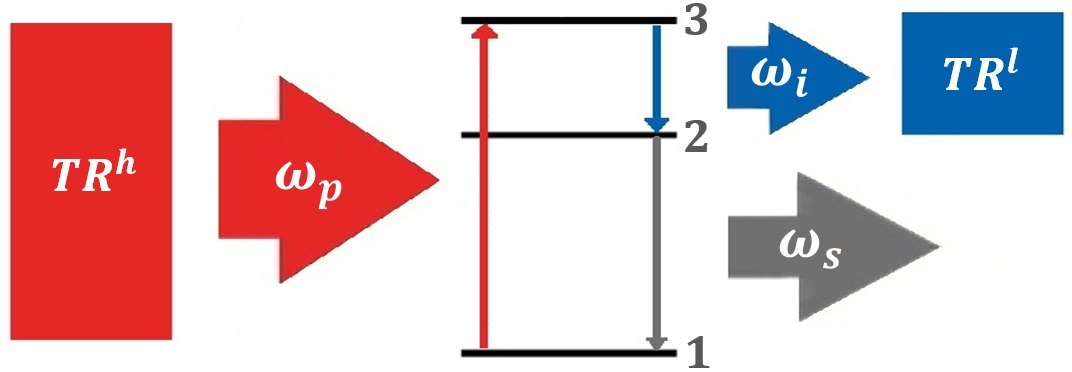}
        \caption{Schematic illustration of a maser or a laser as a QEN, based on Scovil and Schultz-DuBois' three-level quantum design, where the idler mode with frequency $\omega_{i}$ connects to $TR^{l}$, and the pumping process is supported by $TR^{h}$. The signal at frequency $\omega_{s}$, which is the laser light, is interpreted as work generation.}
        \label{fig:3LETlaser}
    \end{figure}

They suggested that a maser could be described as a three-level system that amplifies a signal at frequency $\omega_{s}$, driven by a pump at frequency $\omega_{p}$. The surplus energy is released into an idler mode with frequency $\omega_{i} = \omega_{p} - \omega_{s}$. Scovil and Schultz-DuBois concluded that the maser functions as a thermal engine if the idler mode is connected to $TR^{l}$, while the pumping process is supported by $TR^{h}$, as illustrated in Fig. \ref{fig:3LETlaser}. However, this analysis of masers and lasers remained preliminary and did not evolve into studies exploring solid quantum effects \cite{Myers2022}.

    \begin{figure*}[ht]
        \centering
        \includegraphics[scale=0.32]{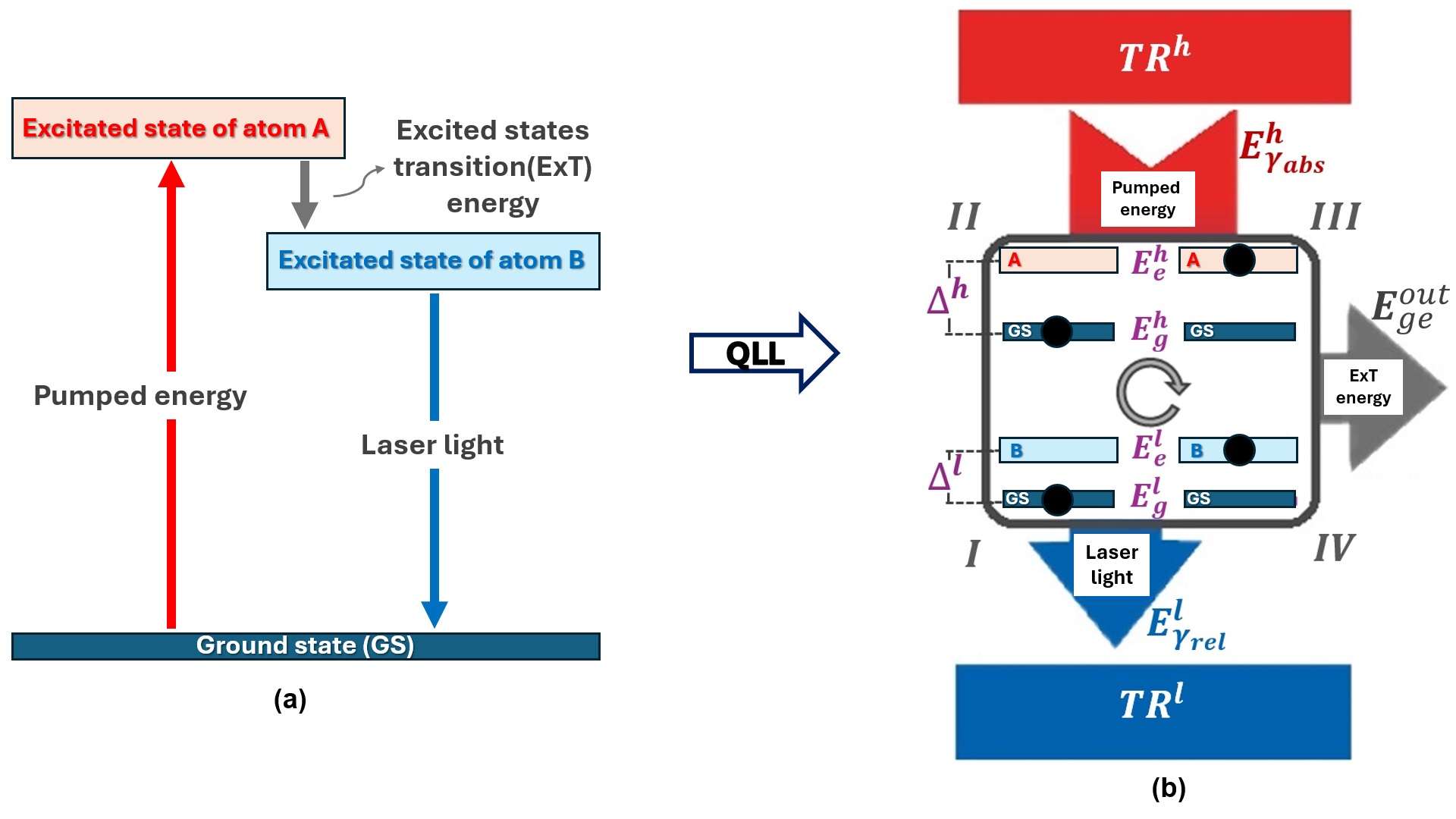}
        \caption{(a) A laser represented as a quantum system with two atoms, A and B, two-levels each. The energy generated in ExT can be understood as $E^{out}_{ge}$ and the laser light as $E^{l}_{\gamma}$ released to the $TR^{l}$. (b) Schematic representation of the laser as a QTM in the OutTransfers'OR.}
        \label{fig:laserQLL}
    \end{figure*}

In this work, we consider the configuration shown in Fig. \ref{fig:laserQLL}a as our laser model. Although it is a simplified and non-generic model, we believe that the analysis presented here is applicable to the other cases as well.

The laser is treated here as a QTM composed of two distinct atoms, each of them separately contributing two energy levels. The electron transition between the two excited states can be undestood as the reason for the change in the dimensional configuration of the laser, where the energy generated during this transition is $E^{out}_{ge}$ and the laser light is the energy $E^{l}_{\gamma}$ released to the $TR^{l}$. Therefore, according to our proposal, the laser can be viewed as a QTM within the OutTransfers'OR, as illustrated in Fig. \ref{fig:laserQLL}b. 

Classifying the laser as a QEN presents some fundamental issues. A QEN must prioritize the generation of energy associated with the expansion of the spatial dimensions of the system (which, classically, would correspond to the performance of mechanical work, $W$). For multilevel systems, this implies reducing the distance between energy levels. Thus, if treated as a QEN, the laser would need to prioritize the generation of idle energy, which is entirely inappropriate. Furthermore, as evidenced in the case of solid-state lasers, the energy released can result from non-radiative relaxation, contradicting the idea that this energy is being released to $TR^{l}$. \cite{Bartolo}

Our proposal is to classify the laser as a QLL, prioritizing the release of energy $E^{l}_{\gamma}$ to the $TR^{l}$, as it should, following all the characteristics proposed in this paper for this QTM design. 

However, the laser does not operate in the Otto cycle, as the occupation probability in this cycle must satisfy the conditions in Eqs. \ref{eq:eq.13} and \ref{eq:eq.14}. As is well known, for a laser to function effectively, population inversion is essential, which can be achieved, for instance, by having two atoms with distinct characteristics. Therefore, it is expected and even desirable that the occupation probability in a laser does not satisfy the conditions of the Otto cycle. As a result, the actual cycle of the laser represents an interesting topic for further research.

\subsection{A Spinless Electron in a One-Dimensional Quantum Ring as the Working Medium of QTMs}

Quantum rings are an important example of quantum system in the physics of low-dimensional materials \cite{Pereira2022,Viefers2004}. Howerver, to our knowledge, nothing has been produced in the QTM context, for quantum rings. Therefore, in this work, we have chosen the $e^-$ in a quantum ring as the two-level quantum system as the working medium. We explore its possible operational regions and the corresponding QTMs operating in the Otto cycle.

According to Viefers et al. \cite{Viefers2004}, the Schrödinger equation for a $e^-$ in a quantum ring depends solely on the polar angle $\varphi$ and can be expressed as

    \begin{equation}
    \mathlarger{-\left(\frac{\hbar ^{2}}{2m_{e}r^{2}} \frac{\partial ^{2}}{\partial \varphi ^{2}}\right)\psi \left( \varphi \right)=E\psi \left( \varphi \right)}, \label{eq:eq.34}
    \end{equation}
where $r$ is the quantum ring radius and $m_{e}$ is the effective mass of the electron.

We take the solutions of Eq. \ref{eq:eq.34} in the form $\psi(\varphi) = e^{im\varphi}$, where $m$ is an integer number. The continuity of the wave function $\psi(\varphi)$ at $\varphi=2\pi$ requires that $m$ be an integer. Using this solution in Eq. \ref{eq:eq.34}, we can find the energy eigenvalues depending on the quantum number $m$. These energies are given explicitly by 

    \begin{equation}
    \mathlarger{E_m = \frac{\hbar^2}{2m_e r^2}m^2}, 
    \label{eq:eq.35}
    \end{equation}
where $m = 1, 2, 3,\ldots$.

Figure \ref{fig:esquema2LET} effectively illustrates the three distinct operational regions of a QTM based on an $e^-$ in a quantum ring, with the understanding that the system's dimensional configuration changes are driven by variations in the radius of the quantum ring.

\begin{table*}[ht]
\centering
\large
\caption{Summary of operational regions of $e^-$ in a quantum ring as the working medium acting between $TR^h$ and $TR^l$ and their corresponding potential QTMs operating in the Otto cycle. Results obtained from the expressions summarized in Table \ref{tab:all_QTMs_table} for $\theta^2=5$ and $\rho = \alpha$, where $\rho=r^{l} / r^{h}$ (Eq. \ref{eq:eq.55}).}
    \includegraphics[scale=0.7]{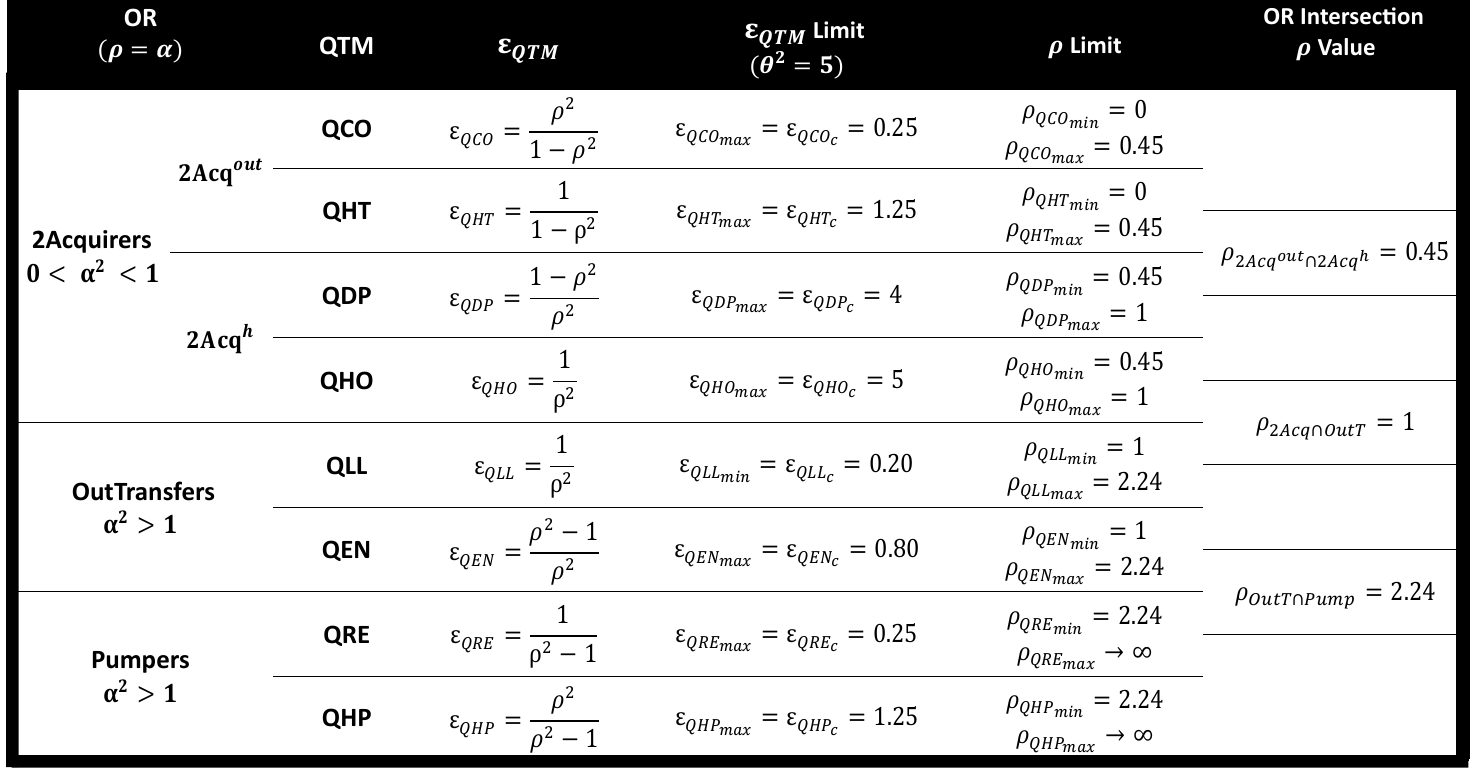} 
    \label{tab:all_QTMs_table_theta=5}
\end{table*}

Since Eq. \ref{eq:eq.35} represents the energy eigenvalues of our quantum system, we can express the energy as

      \begin{equation}
       \begin{split}
             \mathlarger{E^l_g } & \mathlarger{= \frac{\hbar^2}{2m_e {r^l}^2}},   \\
              \mathlarger{E^l_e } & \mathlarger{= 4\frac{\hbar^2}{2m_e {r^l}^2}}  \label{eq:eq.37}
        \end{split}
    \end{equation}
and
      \begin{equation}
       \begin{split}
            \mathlarger{E^h_g } & \mathlarger{= \frac{\hbar^2}{2m_e {r^h}^2}}, \\
             \mathlarger{E^h_e } & \mathlarger{= 4\frac{\hbar^2}{2m_e {r^h}^2}},
            \label{eq:eq.38}
        \end{split}
    \end{equation}
where \( m = 1 \) corresponds to the ground energy level (Eqs. \ref{eq:eq.37}), and \( m = 2 \) corresponds to the excited energy level (Eqs. \ref{eq:eq.38}). Here, \( r^l \) and \( r^h \) denote the quantum ring radius when the energy gaps are \( \Delta^{l} \) and \( \Delta^{h} \), respectively. The ratio \( \Delta^{h}/\Delta^{l} \) can be derived from Eqs. \ref{eq:eq.11} and \ref{eq:eq.16}, and compared with Eq. \ref{eq:eq.44}, which yields

    \begin{equation}
    \mathlarger{\frac{\Delta^{h}}{\Delta^{l}}=\left(\frac{r^{l}}{r^{h}}\right)^2=\alpha^{2}.}   \label{eq:eq.54}
    \end{equation}

Since our system is one-dimensional and changes in its dimensional configuration arise from variations in the radius \( r \), the compression ratio is expected to be defined by the relation

    \begin{equation}
    \mathlarger{\rho = \frac{\text{\normalsize{radius before distancing energy levels}}}{\text{\normalsize{radius after distancing energy levels}}}.}   \label{eq:eq.57}
    \end{equation}

Therefore, we conclude that, for the $e^-$ in a quantum ring acting as the working medium, 

    \begin{equation}
    \mathlarger{\alpha=\rho=\frac{r^{l}}{r^{h}},}   \label{eq:eq.55}
    \end{equation}
where, in accordance with Eq. \ref{eq:eq.44}, we can take \(\left|\alpha\right| = \alpha\).

Eq. \ref{eq:eq.55} can also be used to rewrite the Eqs. \ref{eq:eq.38} as 

  \begin{equation}
       \begin{split}
            \mathlarger{E^h_g } & \mathlarger{=\rho ^2 \frac{\hbar^2}{2m_e {r^l}^2}}, \\
            \mathlarger{E^h_e } & \mathlarger{= 4 \rho ^2 \frac{\hbar^2}{2m_e {r^l}^2}.}  \label{eq:eq.39}
        \end{split}
    \end{equation}

For the calculations and analyses that follow, we use the following experimental parameters: \( T^l = 1K \), \( \theta^2 = 5 \), and \( r^l = 100nm \).

Table \ref{tab:all_QTMs_table_theta=5} presents the results derived from the expressions summarized in Table \ref{tab:all_QTMs_table}, for $e^-$ in a quantum ring as the working medium, functioning between $TR^h$ and $TR^l$ in the Otto cycle.

\begin{figure} [ht]
    \includegraphics[scale=0.29]{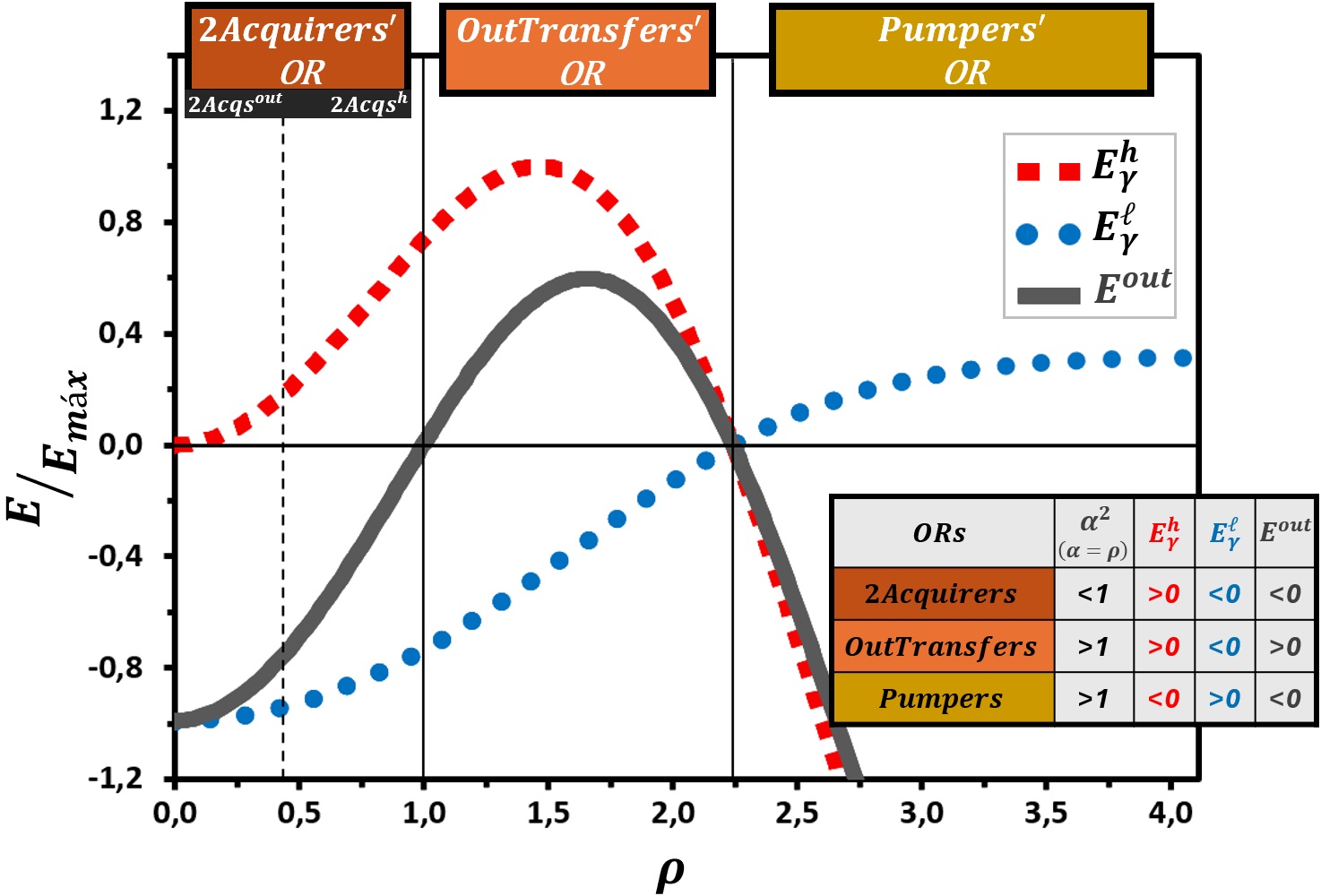}
    \caption{\( E^{h}_{\gamma} \) (red squares), \( E^{l}_{\gamma} \) (blue circles), and \( E^{out} \) (gray solid line) as functions of \( \rho \), normalized by the maximum energy value, for $e^-$ in a quantum ring as the working medium of QTMs operating in the Otto cycle, assuming \( T^l = 1K \), \( \theta^2 = 5 \), and \( r^l = 100nm \). The attached table provides the corresponding energy values, assisting in determining whether the energies are positive—indicating absorption (for TRs) or generation (for QTM's outside)—or negative—indicating release (for TRs) or reception (from QTM's outside). Solid vertical lines mark the boundaries between the operational regions, and the dashed line indicates the subregion division within the 2Acquirers'OR.}
    \label{fig:all_energies_by_rho}
\end{figure}

\begin{figure}[htbp]
    \includegraphics[scale=0.29]{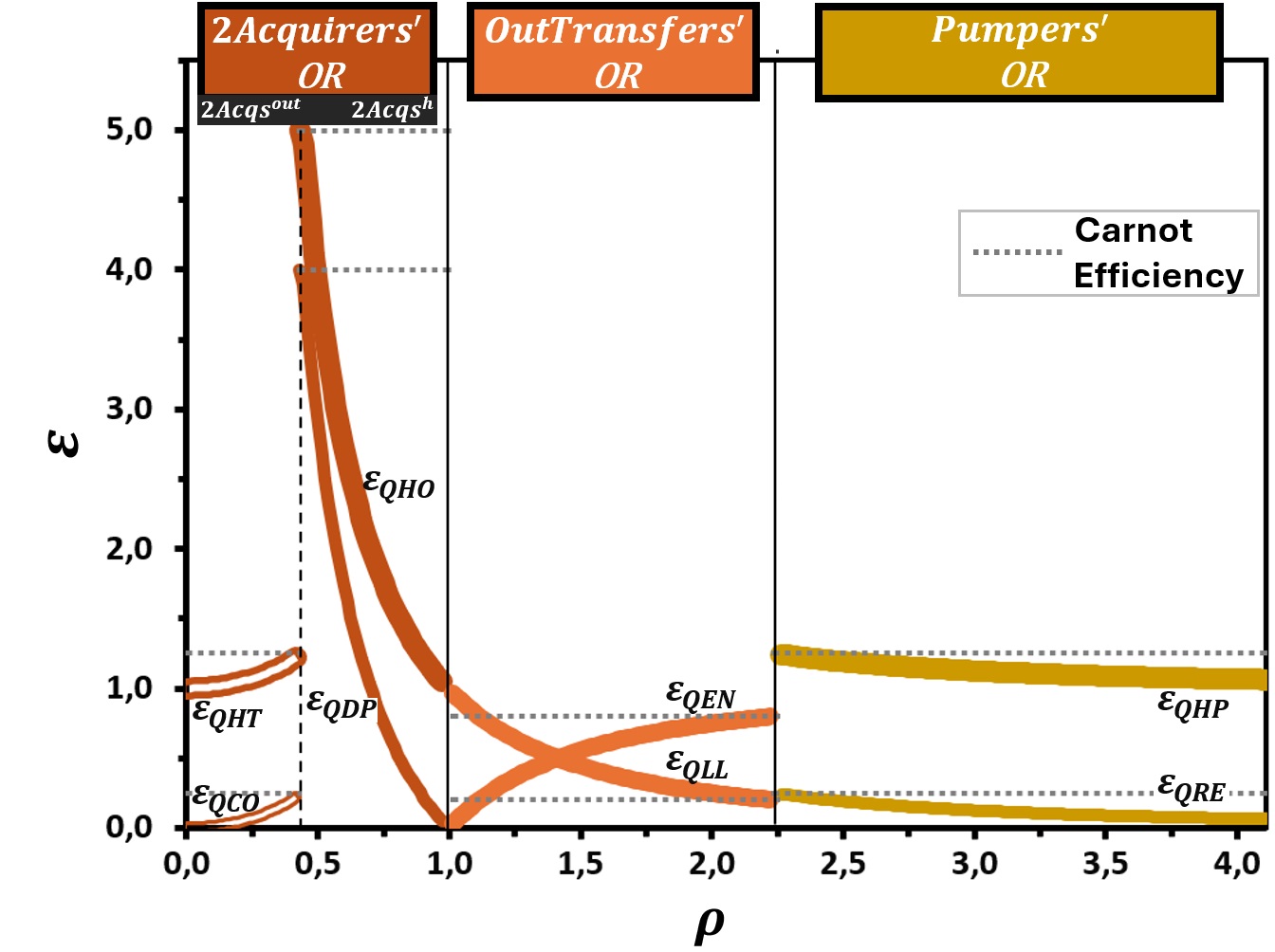}
    \caption{Efficiencies as a function of \( \rho \) for all QTMs based on the $e^-$ in a quantum ring operating in the Otto cycle. The graph was constructed using data from Table \ref{tab:all_QTMs_table_theta=5}. The divisions, indicated by solid vertical lines, and the sub-division, indicated by a dashed vertical line, correspond to those in Fig. \ref{fig:all_energies_by_rho}. The minimum of $\varepsilon_{\textstyle \text{\tiny QLL}}$ and the maximum of any other $\varepsilon_{\textstyle \text{\tiny QTM}}$ occur at the intersection between the efficiency curve of each QTM and the dashed horizontal line, representing the $\varepsilon_{{\textstyle \text{\tiny QTM}}_c}$ of the corresponding QTM. Curve extensions beyond the permitted limits is omitted.}
    \label{fig:all_efficiencies_by_rho}
\end{figure}

The graph in Fig. \ref{fig:all_energies_by_rho} illustrates the energies as a function of the compression ratio \(\rho\), normalized by the maximum energy value, for the three possible operational regions of the system. The dashed lines represent the energies exchanged with the TRs, where blue circles indicate the energy exchanged with \( TR^l \) (\( E^{l}_{\gamma} \)) and red squares represent the energy exchanged with \( TR^h \) (\( E^{h}_{\gamma} \)). The solid gray line denotes the energy exchanged with the QTM's outside (\( E^{out} \)). These energy values are calculated using Eqs. \ref{eq:eq.45}, \ref{eq:eq.46}, and \ref{eq:eq.30}, by substituting the energy levels from Eqs. \ref{eq:eq.37} and \ref{eq:eq.39}. The solid vertical lines divide the graph into three distinct operational regions. The attached table illustrates how these operational regions can be identified by analyzing the $\rho$ values allowed for each region, as well as whether the energies involved are positive, corresponding to absorption processes (for TRs) or energy generation (for QTM's outside), or negative, corresponding to energy release (for TRs) or energy reception (from QTM's outside).

Since \( \alpha = \rho \) (Eq. \ref{eq:eq.55}), the 2Acquirers'OR exists only for \( 0 < \rho < 1 \). Furthermore, we obtain \( E^{h}_{\gamma} > 0 \), \( E^{l}_{\gamma} < 0 \), and \( E^{out} < 0 \) throughout this operational region, indicating that QTMs based on the $e^-$ in a quantum ring can be constructed in the 2Acquirers'OR for any \( \rho \) between 0 and 1. Although it is not visually distinguishable between 2Acq$^{out}$ and 2Acq$^{h}$ subregions, based on Table \ref{tab:all_QTMs_table_theta=5}, we identify the intersection of these subregions at \( \rho = 0.45 \), represented by the dashed vertical line in Fig. \ref{fig:all_energies_by_rho}.

In contrast, the OutTransfers'OR and Pumpers'OR exist only for \( \rho > 1 \). As shown in Fig. \ref{fig:all_energies_by_rho}, for \( \rho \) near 1, \( E^{h}_{\gamma} > 0 \), \( E^{l}_{\gamma} < 0 \), and \( E^{out} > 0 \), corresponding to the OutTransfers'OR. Beyond a certain \( \rho \), these values invert, i.e., \( E^{h}_{\gamma} < 0 \), \( E^{l}_{\gamma} > 0 \), and \( E^{out} < 0 \), indicating the Pumpers'OR. The boundary between these two operational regions is visually clear, and Table \ref{tab:all_QTMs_table_theta=5} shows that this boundary occurs at \( \rho = 2.24 \).

The intersections between 2Acquirers'OR and OutTransfers'OR and between OutTransfers'OR and Pumpers'OR are indicated by the solid vertical lines in Fig. \ref{fig:all_energies_by_rho}.

The efficiencies of all QTMs based on the $e^-$ in a quantum ring operating in the Otto cycle were analyzed as a function of \( \rho \), using the data from Table \ref{tab:all_QTMs_table_theta=5} (Fig. \ref{fig:all_efficiencies_by_rho}). As in Fig. \ref{fig:all_energies_by_rho}, Fig. \ref{fig:all_efficiencies_by_rho} separates the operational regions with solid vertical lines, while dashed lines indicate the subdivisions of subregions. The graph clearly displays the values of $\varepsilon_{\textstyle \text{\tiny QTM}}$ and $\varepsilon_{{\textstyle \text{\tiny QTM}}_c}$ for each QTM within their respective operational regions.

The minimum efficiency for QLL and the maximum efficiencies for all other QTMs are determined at the points where their efficiency curves intersect with the dashed horizontal line, representing the $\varepsilon_{{\textstyle \text{\tiny QTM}}_c}$ of the corresponding QTM. It is important to note that the curves beyond operational limits are omitted to maintain accuracy within the allowable range of \( \rho \).

While a more detailed analysis of these results could be valuable, we believe it is more appropriate for future studies where parameters can be varied and results compared. Nevertheless, we are confident that the consistency and accuracy of the results obtained for this system, summarized in Table \ref{tab:all_QTMs_table_theta=5} and Figures \ref{fig:all_energies_by_rho} and \ref{fig:all_efficiencies_by_rho}, support our innovative approach. These findings suggest that new directions, still at the classical-quantum interface, may be explored to advance and refine studies in Quantum Thermodynamics, particularly concerning QTMs.

\section{Conclusion}

This study introduces an innovative framework for analyzing QTMs by distinguishing between the various ways in which a working medium exchanges energy with \(TR^h\), \(TR^l\), and with the QTM's outside environment. It defines the concept of operational region and proposes new designs of QTMs, including the QDP, QHO, and QLL, which naturally emerge from a detailed analysis of each operational region. In addition to expanding theoretical perspectives, this study introduces a standardized classification for QTMs, providing a cohesive nomenclature and operational framework that we hope will serve as a foundation for future research and practical implementations.

\sloppy{By performing our theoretical analysis for each already known QTM}, we highlight the strong alignment with theoretical predictions, further reinforcing the validity of our findings.

A very significant result is the reclassification of the laser. Instead of being considered a QEN, our proposal categorizes it as a distinct QTM in the OutTransfers'OR. Moreover, the results obtained for two-level quantum systems, particularly for the $e^-$ in a quantum ring as the working medium of QTMs operating in the Otto cycle, highlight the consistency of our approach, since a wide range of possible QTMs was identified, each with its own unique properties and efficiencies.

We believe this work makes notable contributions to Quantum Thermodynamics, expanding theoretical insights while opening new avenues for technological applications. Furthermore, the proposed classification scheme for QTMs offers clarity and cohesion in a field where prior nomenclatures and operational distinctions lacked uniformity. The work developed here is a crucial support for researchers seeking to explore quantum systems with the ability to tune parameters and analyze different operational modes of machines efficiently, without the need to build a new configuration for each adjustment\cite{Gelbwaser2018,jefferson,impurity,PRE1,PRE2}. In this context, it is essential to correctly identify all the operational modes of the machines under study, ensuring that none are overlooked, which enhances the exploration and understanding of their properties and applications. The ideas developed here can also inspire a similar search for nonequilibrium thermodynamic systems\cite{nature}, or also for situations where coherence plays an important role in the first law of thermodynamics\cite{bertulio}, allowing for the investigation of how parameters can be adjusted to explore different states and operational modes. This paves the way for a deeper understanding of the properties and behaviors of these systems. In future studies, we plan to explore parameter variations to optimize the proposed QTMs further, enhancing the link between theory and practical implementation in quantum devices.
\section*{Funding}
{This work was partially supported
by the Brazilian agencies CNPq and FAPEMIG: C. Filgueiras and M. Rojas acknowledge FAPEMIG Grant No. APQ 02226/22. C. Filgueiras acknowledges CNPq Grant No. 310723/2021-3 and M. Rojas acknowledges CNPq Grant No 317324/2021-7.}
\section*{Data Availability Statement}
No new data were created or analysed in this study. Data sharing is not applicable to this article.
\section*{Conflicts of Interest}
The authors declare no conflicts of interest


\begin{thebibliography}{9}
\bibitem{Springer2018} F. Binder, L. A. Correa, C. Gogolin, J. Anders, and G. Adesso, eds., \textit{Thermodynamics in the Quantum Regime: Fundamental Aspects and New Directions}, 1st ed. (Springer, Cham, 2018).

\bibitem{Dann2023} R. Dann and R. Kosloff, New J. Phys. \textbf{25}, 043019 (2023).

\bibitem{Adlam2022} E. Adlam, L. Uribarri, and N. Allen, AVS Quantum Sci. \textbf{4}, 022001 (2022).

\bibitem{Aw2021} C. C. Aw, F. Buscemi, and V. Scarani, AVS Quantum Sci. \textbf{3}, 045601 (2021).

\bibitem{Goold2021} J. Goold and K. Modi, AVS Quantum Sci. \textbf{3}, 045001 (2021).

\bibitem{Trushechkin2022} A. S. Trushechkin, M. Merkli, J. D. Cresser, and J. Anders, AVS Quantum Sci. \textbf{4}, 012301 (2022).

\bibitem{CleversonRef} C. Filgueiras \textit{et al.}, Int. J. Geom. Methods Mod. Phys. \textbf{20}(14), 2450009 (2023).

\bibitem{Myers2022} N. M. Myers \textit{et al.}, AVS Quantum Sci. \textbf{4}, 027101 (2022).

\bibitem{Scovil1959} H. E. D. Scovil and E. O. Schulz-DuBois, Phys. Rev. Lett. \textbf{2}, 262 (1959).

\bibitem{Kieu2004} T. D. Kieu, Phys. Rev. Lett. \textbf{93}, 140403 (2004).

\bibitem{Quan2007} H. T. Quan \textit{et al.}, Phys. Rev. E \textbf{76}, 031105 (2007).

\bibitem{Gelbwaser2018} D. Gelbwaser-Klimovsky \textit{et al.}, Phys. Rev. Lett. \textbf{120}, 170601 (2018).

\bibitem{Zheng2014} Y. Zheng and D. Poletti, Phys. Rev. E \textbf{90}, 012145 (2014).

\bibitem{Torrontegui2017} E. Torrontegui \textit{et al.}, Phys. Rev. A \textbf{96}, 022133 (2017).

\bibitem{Uzdin2014} R. Uzdin and R. Kosloff, New J. Phys. \textbf{16}, 095003 (2014).

\bibitem{Huang2013} X. L. Huang, H. Xu, X. Y. Niu, and Y. D. Fu, Phys. Scr. \textbf{88}, 065008 (2013).

\bibitem{Wang2007} J. Wang, J. He, and Z. Mao, Sci. China Phys. Mech. Astron. \textbf{50}, 163 (2007).

\bibitem{Lin2003} B. Lin and J. Chen, J. Appl. Phys. \textbf{94}, 6185 (2003).

\bibitem{Filgueiras2019} C. Filgueiras, Results Phys. \textbf{15}, 102556 (2019).

\bibitem{Khlifi2020} Y. Khlifi, A. El Allati, A. Salah, and Y. Hassouni, Int. J. Mod. Phys. B \textbf{34}, 2050212 (2020).

\bibitem{ElHawary2023} K. El Hawary and M. El Baz, Quantum Inf. Process. \textbf{22}, 190 (2023).

\bibitem{Kosloff2017} R. Kosloff and Y. Rezek, Entropy \textbf{19}, 136 (2017).

\bibitem{Pereira2022} L. F. C. Fernando, M. M. Cunha, and E. O. Silva, Body Syst. \textbf{63}, 58 (2022).

\bibitem{Viefers2004} S. Viefers, P. Koskinen, P. S. Deo, and M. Manninen, Physica E \textbf{21}(1), 1 (2004).

\bibitem{Bartolo} B. Di Bartolo, Optical Interactions in Solids (Wiley, New York, 1991).

\bibitem{jefferson} J. L. Diniz de Oliveira, M. Rojas, and C. Filgueiras, Phys. Rev. E \textbf{104}, 014149 (2021).

\bibitem{impurity} A. Prakash, A. Kumar, and C. Benjamin, Phys. Rev. E \textbf{106}, 054112 (2022).

\bibitem{PRE1} R. J. de Assis, J. S. Sales, J. A. R. da Cunha, and N. G. de Almeida, Phys. Rev. E \textbf{102}, 032111 (2020).

\bibitem{PRE2} S. Chand and A. Biswas, Phys. Rev. E \textbf{95}, 032111 (2017).

\bibitem{nature} M. M. Ali, W. M. Huang, and W. M. Zhang, Sci. Rep. \textbf{10}, 13500 (2020).

\bibitem{bertulio} B. L. Bernardo, Phys. Rev. E \textbf{102}, 062152 (2020).

\end{thebibliography}
\end{document}